\documentclass[11pt,a4paper]{article}
\pdfoutput=1
\usepackage{jheppub}
\usepackage{amsmath,amscd}
\usepackage{amsfonts}
\usepackage{amssymb}
\usepackage{graphicx}
\usepackage{color}
\usepackage[normalem]{ulem}
\usepackage{hyperref}
\usepackage{mathtools}

\newcommand{\CC}{\mathbb{C}} % Complessi
\newcommand{\RR}{\mathbb{R}} % Reali
\newcommand{\ZZ}{\mathbb{Z}} % Interi
\newcommand{\G}{\mathcal{G}}

\def\tr         {{\rm  tr}}

\def\calf         {{\cal F}}

\def\call         {{\cal L}}
\def\calm         {{\cal M}}

\def\calo         {{\cal O}}

\def\cals         {{\cal S}}
\def\calt         {{\cal T}}

\def\be{\begin{equation}}
\def\ee{\end{equation}}
\def\bea{\begin{eqnarray}}
\def\eea{\end{eqnarray}}

\def\a{\alpha}
\def\b{\beta}
\def\h{\eta}

\def\G{\Gamma}
\def\d{\delta}
\def\e{\epsilon}
\def\D{\Delta}
\def\l{\lambda}
\def\L{\Lambda}

\def\f{\phi}

\def\m{\mu}
\def\n{\nu}
\def\o{\omega}
\def\O{\Omega}
\def\p{\pi}
\def\r{\rho}

\def\s{\sigma}
\def\S{\Sigma}
\def\t{\tau}

\def\sF{{{ F}\!\!\!\!\hskip.8pt\hbox{\raise1pt\hbox{/}}\,}}
\def\som{{{ \omega}\!\!\!\!\hskip.8pt\hbox{\raise1pt\hbox{/}}\,}}
\def\sJ{{{\rm J}\!\!\!\!\hskip.8pt\hbox{\raise1pt\hbox{/}}\,}}

\def\F{\Phi}
\def\pa{\partial}

\def\to{\rightarrow}
\def\nonu{\nonumber \\{}}
\def\half{{1 \over 2}}

\title{Holography for  bulk states in 3D quantum gravity}

\author{Joris Raeymaekers and Gideon Vos}

\affiliation{CEICO, Institute of Physics of the ASCR,  Na Slovance 2, 182 21 Prague 8, Czech Republic.}
\emailAdd{joris@fzu.cz}
\emailAdd{vos@fzu.cz}

\abstract{In this work we discuss the holographic description of states in the Hilbert space of (2+1)-dimensional quantum gravity, living on a time slice in the bulk. We focus on pure gravity coupled to pointlike sources for heavy spinning particles.  We develop a formulation where the  equations for the backreacted metric reduce to two decoupled  Liouville equations with delta-function sources under pseudosphere boundary conditions. We show that both the semiclassical wavefunction  and the  gravity solution are determined by a universal object, namely a classical Virasoro vacuum block on the sphere. In doing so we derive a version of Polyakov's conjecture,  as well as an existence criterion, for classical  Liouville theory  on the pseudosphere.
We also discuss how some of these results are modified when considering closed universes with compact spatial slices.}
\arxivnumber{}
\keywords{}
\begin{document}
 \maketitle

\section{Introduction and summary} 
In its most common formulation, the AdS/CFT correspondence \cite{Maldacena:1997re}
allows for the computation of   correlation functions in a strongly coupled CFT  from   AdS gravity in the bulk with sources placed on the boundary \cite{Gubser:1998bc, Witten:1998qj}. Though the CFT description is believed to capture the full gravitational Hilbert space, it is far from obvious how quantum gravity states, defined on some  time slice $\S$, are described in CFT terms. It would  therefore be desirable to have a more direct holographic dictionary   relating  quantum states in the bulk to CFT quantities. Such a reformulation of holography was referred to as `CFT/AdS' by H. Verlinde in an inspiring lecture \cite{VerlindeCFTAdS}. It would allow one to  address
 important issues such as  the description of local excitations or of the black hole interior. These are  closely related to the question of holographic emergence of  bulk locality   which is a subject of intensive research. 
 
 In this work  we will address the holographic description of certain bulk states   in 2+1 dimensional quantum gravity. 
 Our scope will be rather modest, focusing  on states containing  a number of   point-like particles but no black holes.  Though we will make heavy use of the quasi-topological nature of (2+1)D gravity and of its Chern-Simons description, one might hope that some of our  results, such as the connection  to CFT conformal blocks that we will uncover, generalize to higher dimensions as well. A different approach to bulk state holography was proposed in \cite{Araujo-Regado:2022gvw}.

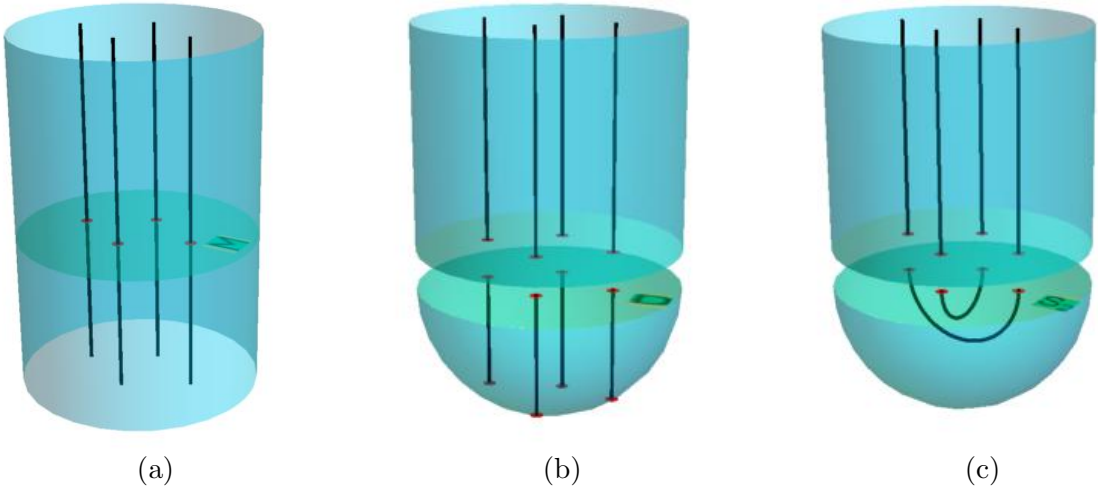
\begin{figure}
%	\begin{center}
		\begin{picture}(300,100)
\put(-20,5){\includegraphics[height=180pt]{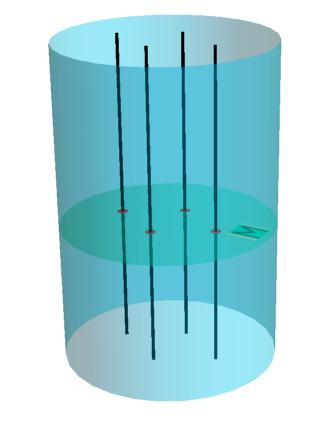}}
	\put(110,0){\includegraphics[height=180pt]{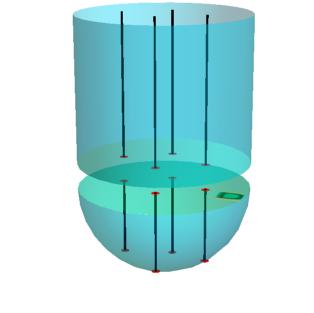}}
		\put(260,0){\includegraphics[height=180pt]{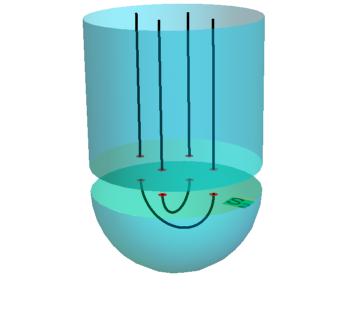}}
		\put(50,0){(a)}\put(202,0){(b)}\put(360,0){(c)}
		\end{picture}
%		\end{center}
		\caption{(a) A collection of point particles in AdS piercing an initial time slice $\S$. (b) When $\S$ is a disk, we prepare the state with a path integral over  3-geometries  with particles emerging from the boundary in the past. (c) When $\S$ is a two-sphere, we compute a path integral over 3-geometries without past boundary. }
	\label{FigHH}
\end{figure}

More precisely, we will consider  states in the bulk which contain 
 a number of massive, spinning particles in AdS in  a point-like limit. Their worldlines pierce an initial time slice $\S_0$ with the topology of a disk, as sketched schematically in Figure \ref{FigHH}(a). 
 
The particles are taken to be sufficiently heavy to backreact on the geometry and produce a  multi-centered AdS solution. 

  At the classical level, the state of the gravitational field we are interested in consists of initial data on $\S_0$ which satisfy the Hamiltonian and momentum constraints, and  has the appropriate behavior near the particle sources. We will introduce a parametrization of (2+1)D gravity where these initial value constraints reduce to two decoupled Liouville equations with delta-function sources
 \be \pa_z \pa_{\bar z} \F+ e^{-2 \F} = \p \sum_i \a_i \d^{(2)}(z - z_i), \qquad \pa_z \pa_{\bar z} \tilde \F+ e^{-2\tilde \F} = \p \sum_i \tilde \a_i \d^{(2)}(z - z_i)\label{Liouvintr}
\ee
Here, complex coordinates range over the upper half plane, with $z_i$  the locations where the particle worldlines pierce the initial slice $\S_0$, and 
$\a_i = 2 G (m_i + s_i),\ \tilde \a_i = 2 G (m_i - s_i)$, with $m_i$ and $s_i$ the particle masses and helicities respectively. The Liouville fields $\F$ and $\tilde \F$ obey the  pseudosphere boundary conditions of \cite{Zamolodchikov:2001ah} on the real line.
 The metric on the initial value surface is then given in terms of these fields as
\be 
ds^2_{\S_0} =\left( e^{- \F}  +  e^{-  \tilde  \F}  \right)^2 dz d\bar z -  \left( {\rm Im} \left( \pa_z (\F -\tilde \F) dz \right)\right)^2. 
\ee
To find the full (2+1)D metric one needs to integrate a system of first order  time evolution % of the metric can be reduced to a system of first order 
equations spelled out in (\ref{dyneqs}) below. While generically quite complicated, the time dependence simplifies in  cases where all the particles are either `chiral' or `antichiral', meaning that $\a_i \tilde \a_i=0, \ \forall i$. The full 2+1 dimensional metric is in that case  
\be
ds^2 
= \left| e^{- \F(z_+, \bar z_+)} d z_+   +  e^{-  \tilde  \F(z_-, \bar z_-)} d z_- \right|^2 -  \left({\rm Im} \left( \pa_{z_+} \F(z_+, \bar z_+) d z_+ - \pa_{z_-}\tilde\F(z_-, \bar z_-) d z_-\right)\right)^2 , \nonu
\ee
where $z_\pm = z \pm t$.
This class of metrics  generalizes that studied in \cite{Hulik:2016ifr}, where purely  chiral solutions with $\tilde \a_i =0, \forall i$ were considered. 

On the quantum level, a useful way to represent a bulk state is as a Hartle-Hawking path integral over three-surfaces with the time slice $\S_0$ as boundary.  Our goal is  to give a direct CFT interpretation of this object.   To  prepare the multi-particle state on $\S_0$ we perform a path integral  on a 3-manifold $X$ ending on $\S_0$,  containing particle worldlines emerging from the boundary in the past and terminating on
 $\S_0$  
 (see Figure \ref{FigHH}(b)). In the leading  approximation, we have to evaluate the classical on-shell action on $X$ with appropriate boundary terms. %The conclusion in both cases is that, in the leading classical approximation, the Hartle-Hawking wavefunction reduces to the exponential of the on-shell Liouville action,
In doing so we will find a close connection between the wavefunction and a holomorphic Virasoro conformal block on the sphere. We will find a relation of the form
\be 
\Psi_{HH} [(z_i,\bar z_i), \a_i, \tilde \a_i ] \sim N(\a_i, \tilde \a_i) \exp\left({ -{k \over 4} ( F_0 ( z_i,  \bar z_i; \a_i) +  F_0 ( z_i,\bar z_i; \tilde \a_i) }\right)\label{Psiintr}
\ee
Here  $N(\a_i, \tilde \a_i)$ 
is  a normalization factor independent of the $z_i$, and
$ F_0 ( z_i,  \bar z_i; \a_i)$ is a classical  Virasoro   block for a $2n$-point  correlator on the sphere, with identical operators inserted in each $z_i$ and its image 
point $\bar z_i$. The subscript in $F_0$ indicates that it is  a vacuum block in  the channel  where the operators in image points  are fused pairwise, as illustrated in Figure \ref{fig:OPEchannel}. The fact that, in the semiclassical approximation, the  features of the state are captured by a universal object insensitive to the details of the UV completion, is a concrete realization of general expectations  \cite{Fitzpatrick:2014vua}. In the process of deriving the relation (\ref{Psiintr}), we will also derive a version of Polyakov's conjecture \cite{Zograf_1988,Takhtajan:2001uj} for Liouville theory on the pseudosphere, and give an existence criterion for Liouville solutions on the pseudosphere in terms of a certain reflection property of the classical block $F_0$.

The classical block $F_0$ does  not only determine the Hartle-Hawking wavefunction, but also the full 2+1 dimensional metric. The Liouville solution in (\ref{Liouvintr}) is of the form
\be 
e^{\F} = i (\psi_1 \bar \psi_2- \bar \psi_1 \psi_2 )
\ee
where 
$\psi_{1}(z) $  and $\psi_{2}(z) $ are the two independent  solutions (with unit Wronskian, {i.e. $\psi_1'\psi_2-\psi_1\psi_2'=1$}) to an ordinary differential equation whose coefficients are determined by $F_0$, namely
\be 
(\pa_z^2 + T(z) )\psi  =0,
\ee 
with
\be 
T (z) =\sum_{i=1}^{n}  \left(   {\a_i (2- \a_i)   \over 4  (z-z_i)^2}+ { \a_i (2- \a_i)  \over  4 (z-\bar z_i)^2} + {\pa_{z_i} F_0 \over z-z_i}+ {\overline{\pa_{z_i} F_0}\over z-\bar z_i}\right).\label{Tintr}
\ee

Our approach extends in principle  to initial slices with a different topology.  It is interesting to consider the case where $\S_0$ is compact without boundary. Here  we expect to find a CFT description of  closed universes with negative cosmological constant which contain some point  particles. We will focus on the case where  $\S_0$ has spherical  topology,
and consider a  Hartle-Hawking path integral   over a ball containing worldlines of particles  pair-created in the past (see Figure \ref{FigHH}(c)). In contrast  to the case with conformal boundary, the wavefunction and backreacted gravity solution now depend on dynamical CFT data, rather than on conformal kinematics alone.  Indeed, we will find  relations similar to (\ref{Tintr}) and (\ref{Psiintr}) above, where the vacuum block $F_0$ is replaced by a specific non-vacuum block
\be 
F( z_i, \a_i, \b^*_I ),
\ee
 where $\b^*_I$ label the exchanged conformal families.  These depend implicitly on the insertion points, $\b^*_I = \b_I^*(z_i, \bar z_i)$, in a highly complicated manner. In fact, to determine them one needs  dynamical information in the form of in the large $c$ behavior of the Liouville three-point  functions. 
 
This paper is organized as follows. In Section \ref{Sechyp} we introduce a  convenient parametrization of (2+1)D gravity in terms of two auxiliary 2D hyperbolic metrics. A judicious choice of gauge  allows us to rephrase the  backreaction  of spinning point particles as  a pair of decoupled inhomogeneous Liouville equations. In Section \ref{Secasads} we apply our formalism to asymptotically AdS spacetimes and derive a connection between  the  Hartle-Hawking-like wavefunction, the on-shell Liouville action on a pseudosphere, and a classical Virasoro block on the sphere. For this purpose we derive a version of Polyakov's conjecture and give a new existence criterion for Liouville theory on the pseudosphere.  We also discuss some  special cases and explicit examples. In Section \ref{SecS2} we discuss the modifications occurring when considering closed universes with compact spatial slices.
We end with some open problems and future directions.

\section{`Doubly hyperbolic' parametrization of 2+1 dimensional gravity}\label{Sechyp}
In this section we introduce a convenient parametrization of (2+1)D gravity.
%description of 3D gravity which will facilitate our inverstigations. 
It relies heavily on the formulation of the theory as a Chern-Simons
theory with gauge group $SL(2,\RR) \times SL(2,\RR)$. The  constraint equation of $SL(2,\RR)$ Chern-Simons theory, which states that the field strength should vanish on spacelike slices $\S$, can be interpreted as the 2D Euclidean Einstein equation with negative cosmological
constant \cite{Verlinde:1989ua}. In this way we obtain a reformulation of (2+1)D gravity as a time evolution of two auxiliary hyperbolic metrics.
Perhaps unsurprisingly, the  problem of  of backreacting point-particle sources then reduces to solving the Liouville equation on $\S$ with delta-function sources.

\subsection{Chern-Simons description }

Our starting point is the Chern-Simons formulation of (2+1)D gravity:
\be
S = S_{CS} [A]-   S_{CS} [\tilde A], \qquad 
 S_{CS} [A] = {k \over 4\p} \tr  \int_\calm \left( A  \wedge dA + {2 \over 3} A \wedge A \wedge A \right), \label{CSaction}
\ee
where $A$ and $\tilde A$ are  potentials for the gauge group $SL(2,\RR)$.   The gauge potentials are related to the vielbein $E$ and spin connection $\O$ as\footnote{Throughout, we work in units where the AdS radius is set to one.}
\be 
A = \O + E, \qquad \tilde A = \O - E.\label{EOmega}
\ee
The generators of $sl(2,\RR)$ satisfy the commutation relations $[L_m,L_n]= (m-n) L_{m+n}$,  $m, n  =0, \pm 1$.  The trace in (\ref{CSaction}) will, for definiteness, be taken in the 2-dimensional representation\footnote{For concreteness we can take
\be
L_0 = \half \left(\begin{array}{cc} 1 &0\\0 &-1\end{array} \right),\qquad L_1=  \left(\begin{array}{cc} 0 &-1\\0 &0\end{array} \right),  \qquad L_{-1} =  \left(\begin{array}{cc} 0 &0\\1 &0\end{array} \right).\ee }, where $\tr L_a L_b =\half \h_{ab}$.  The Chern-Simons level $k$ is then related to the Brown-Henneaux central charge \cite{Brown:1986nw} as
\be 
c := {3 \over 2 G} = 6 k.
\ee

In the present Section take the manifold $\calm$ on which the Chern-Simons theory is defined to be of the form $\RR \times \S$, where  the real line is the time direction and $\S$ is a Riemann surface. 
For concreteness we will use $t \in \RR$ as the time coordinate and choose a local complex coordinate $z$ on $\S$. To clarify the canonical structure, it is useful to denote  the spatial projections of $A, \tilde A$ by  a hat,
\be 
\hat A \equiv A_z dz + A_{\bar z} d\bar z,\qquad \hat {\tilde A} \equiv \tilde A_z dz + \tilde A_{\bar z} d\bar z
\ee
and similarly introduce a spatial  exterior derivative 
\be \hat d = d z \pa_z + d\bar z \pa_{\bar z}.\ee
The canonical structure of the Chern-Simons action is clarified by rewriting it, up to a total derivative, as
 \be
  S[A]  = - {k \over 4\p} \tr \int_{\RR \times \S} dt \wedge \left(  \hat A  \wedge\dot{\hat A} - 2 A_t \hat F \right).\label{Scandec}
 \ee
The time component $A_t$ is not a dynamical variable but a Lagrange multiplier enforcing the {\em constraint equation} 
\be 
\hat F = \hat d\hat A +\hat A \wedge \hat A =0 \label{constrF},%, \qquad \tilde{\hat F} = d\hat {\tilde A} + \hat {\tilde A} \wedge \hat {\tilde A} =0,
\ee
The remaining (complex) equation of motion
\be 
F_{tz} =0\label{dynF}
\ee
is a {\em dynamical equation} describing the time evolution of the dynamical variables $\hat A$. 

\subsection{Choice of gauge}

In what follows, we will use the gauge freedom to work in a generalization of the temporal gauge $A_t=\tilde A_t =0$.  It involves choosing two smooth vector fields $V$ and $\tilde V$ which are tangent to $\S$ at all times, i.e.
\be
V = V^z \pa_z +V^{\bar z} \pa_{\bar z}, \qquad \tilde V = \tilde V^z \pa_z +\tilde V^{\bar z} \pa_{\bar z}.
\ee
We then impose the gauge conditions
\be
A_t =- i_V {\hat A} , \qquad 
\tilde A_t =- i_{\tilde V} \tilde{\hat  A}   \label{covgauge}.
\ee
In the nomenclature of \cite{Henneaux:1992ig}, this is a Lagrange multiplier gauge which, for  $V=\tilde V=0$, reduces to the temporal gauge.
We will keep the vector fields $V$ and $\tilde V$ general for the moment; in what follows  they will be   chosen judiciously in order to simplify the problem. We should note that, if  $V$ and $\tilde V$  coincide on some locus, then the vielbein $E$  (see (\ref{EOmega})) degenerates there and we have a coordinate singularity. Below, when adding particle sources, this will actually happen on the isolated worldlines, reflecting the conical singularity  of  the metric there \cite{Deser:1983tn,Deser:1983nh}. % For similar considerations see section 4.2.1 of \cite{Denef:2000nb} 

Using (\ref{covgauge}) to eliminate $A_t$, the equations for the dynamical variables reduce to 
\begin{align}
\hat F &= 0, & \hat {\tilde F} &= 0\label{eom1}\\
\pa_t {\hat A} &= - \hat \call_V \hat A, & \pa_t {\hat{\tilde A}} &= - \hat \call_{\tilde V} \hat {\tilde A}\label{eom2}
\end{align}
The first line contains the constraint equations on the time slice $\S$, while on the second line we have the dynamical equations determining the time evolution.

The gauge choice  (\ref{covgauge}) has residual  symmetries stemming from parameters satisfying\footnote{Here and in what follows, we define the spatial Lie derivative $\hat \call_V = i_V \hat d + \hat d i_V$m to act on form indices only, and not on  Lie algebra indices.}
\be 
\dot \L = - \hat \call_V \L , \qquad \dot{\tilde \L} = - \hat \call_{\tilde V} \tilde \L.\label{resgauge}
\ee
The parameter $\L$ can be chosen arbitrarily  on specific time slice, say at $t=0$, 
which can be used to bring $\hat A, \tilde{\hat A}$ to a convenient form there. 

\subsection{`Doubly hyperbolic' variables}

We now introduce a convenient parametrization for the spatial  connections $\hat A$ and $\hat{\tilde A}$.  This parametrization naturally arises from the relation \cite{Verlinde:1989ua} between flat $SL(2,\RR)$ connections  and constant negative  curvature metrics on $\S$.

Concretely, we parametrize $\hat A, \hat{\tilde A}$ as
\be 
\hat A =  e L_{1} -  \bar e L_{-1} + i \o L_0, \qquad \hat{\tilde A} = - \tilde e L_{1} +  \bar{\tilde e} L_{-1} +i \tilde  \o L_0,\label{gaugepot2D}
\ee
where the one-forms $e$ and $\bar e$ are related by complex conjugation and $\o$ is real\footnote{Note that with these reality conditions $\hat A$ and $\hat{\tilde A}$
actually take values in $su(1,1)$. A similarity transformation with $e^{i \p (L_1 - L_{-1})/4}$ would yield $sl(2,\RR)$-valued potentials, though  here we will stick with the simpler form (\ref{gaugepot2D}).\label{FNsu11}}, and similarly for the quantities with a tilde.
 The constraint equations (\ref{eom1})
reduce to
\bea
\hat d e - i \o \wedge e =&0,\label{vielbpost}\\
%\hat d \bar e + i \o \wedge \bar e =&0,\\
\hat d \o + 2i e \wedge \bar e   =&0,\label{Fzero}
\eea
and similarly for $\tilde e, \tilde \o$.
%\begin{align}
%\hat d e - i \o \wedge e =&0, & \hat d \tilde e- i \tilde  \o \wedge \tilde e =&0\\
%\hat d \bar e + i \o \wedge \bar e =&0, & \hat d \bar{\tilde e} + i \tilde \o \wedge \bar{ \tilde e} =&0\\
%\hat d \o + 2i e \wedge \bar e   =&0,& \hat d \tilde  \o + 2 i \tilde e \wedge \bar{ \tilde e}   =&0.\label{Fzero}
%\end{align}
%\comment{Compare signs with 2d einstein eqs.}
These equations state that, at each time $t$, the two auxiliary metrics on $\S$ 
\be 
ds^2_2 = e \bar e,\qquad  d\tilde s^2_2 =\tilde  e \bar{\tilde  e}\label{auxmetrs}
\ee
have constant negative curvature,
while  $\o , \tilde \o $  the role of the associated spin connection one-forms. The latter  are determined algebraically in terms of $e, \tilde e$ and their spatial derivatives through (\ref{vielbpost}). 

The dynamical equations (\ref{eom2}) become
\be 
\dot e = - \hat \call_V e, \qquad \dot {\tilde e} = - \hat \call_{\tilde V} \tilde e.\label{dyneqse}
\ee
The dynamical equation for the spin connection, $\dot \o = - \hat \call_V \o$ is automatically satisfied when expressing $\o$ in terms of $e$ and using (\ref{dyneqse}).

The actual (2+1)D gravity metric $ds^2 = \tr (A - \tilde A)^2/2$ takes the form
\be 
ds^2 = \left|\left(  i_V e +   i_{\tilde V} {\tilde e} \right) dt -e -\tilde e\right|^2
-{1 \over 4} \left( \left( i_V \o -  i_{\tilde V} {\tilde \o} \right) dt -\o +\tilde \o \right)^ 2.\label{3Dmetrgen}
\ee
It is generically related to  the two auxiliary constant curvature metrics (\ref{auxmetrs}) in a nontrivial way.
In the special left-right symmetric case where $ e= \tilde e $, the metric on $\S$
 is simply proportional to the auxiliary constant negative curvature metrics  (\ref{auxmetrs}), $ds^2_\S = 4 ds^2_2 = 4 d\tilde s^2_2$. Furthermore, if $V = - \tilde V$, the (2+1)D metric in this class is static. In Section (\ref{Secexs}) we will discuss another class of solutions, obtained as a certain scaling limit, which are fibrations over a 2D base with hyperbolic metric $ds^2_2$.

\subsubsection*{More on the phase space}

While the  parametrization of the gauge potentials (\ref{gaugepot2D},\ref{vielb},\ref{mudisk})  may seem cumbersome from the point of view of  (2+1)D metric variables,  it gives a useful `left-right factorized' description of the gravity phase space which, following \cite{Verlinde:1989hv},  facilitates the connection to 2D CFT. Indeed, from the above parametrization we see that the space of flat $SL(2,\RR)$ gauge potentials (\ref{gaugepot2D}) for which the 2D vielbein $(e, \bar e)$ is invertible\footnote{As shown in \cite{Kim:2015qoa},\cite{Eberhardt:2022wlc}, the full moduli space of flat  $SL(2,\RR)$ connections consists of several components, one of which is the Teichm\"uller space and which is picked out by the restriction to inverible vielbeins.}, modulo gauge transformations, is given by the space of constant negative curvature  vielbeins  modulo diffeomorphisms and local Lorentz transformations. In other words, this phase space is, schematically,
\be 
%{{\rm constant\ negative\ curvature\ zweibeins\ on\ \S} \over {\rm Diff_0 (\S)\ \times Local\ Lorentz }}\simeq
{{\rm metrics\ on\ \S} \over {\rm Diff_0 (\S)\ \times Weyl }} .\label{phaseteich}
\ee
If $\S$ is of genus $g$ and has $b$ boundaries, this  space can be identified as the Teichm\"uller space $\calt (g, b )$. From the 2+1 decomposition of the the Chern-Simons action  (\ref{Scandec}) we see that the symplectic form on the phase space is given by
\be 
\o_{CS} = - { k \over 4 \p} \tr \int_\S \d \hat A \wedge \d \hat A ,
\ee
which can be shown to be equivalent to the standard  Weil-Petersson symplectic form on $\calt (g,b)$ \cite{Verlinde:1989ua}.
Combining with the  analogous $\hat{\tilde A}$ phase space we conclude that the phase space of (2+1)D gravity  is, locally\footnote{While (\ref{phasespaceTT}) is a correct  correct local description of the phase space of (2+1)D gravity, it was argued in \cite{Kim:2015qoa, Eberhardt:2022wlc} that invariance under large diffeomorphisms requires a further  quotient by the diagonal action of the mapping class group $M$, which leads to
${\calt (g,b) \times \calt (g,b)\over M}.$
Such global aspects will however not play a role in the semiclassical considerations in this work. \label{FNMappingclass}},
\be 
\calt (g,b) \times \calt (g,b).\label{phasespaceTT}
\ee
A different parametrization of the field space, which is more natural in metric variables, describes the phase space as the cotangent bundle of a single copy of the Teichm\"uller space, $T^* \left(\calt (g,b) \right)$. These two descriptions are  equivalent in a  nontrivial way, as shown in \cite{Mess,Scarinci:2011np}.

 We should remark that, if the boundaries are asymptotic boundaries, which will be the case of interest for us, the diffeomorphisms ${\rm Diff_0 (\S)}$ in (\ref{phaseteich}) are defined to act trivially at infinity, and the resulting 
Teichm\"uller space is in fact infinite dimensional \cite{Verlinde:1989hv}. It captures  `boundary graviton' excitations which we will describe more explicitly  in Section \ref{Secasads}. For example, for $g=0, b=1$, it can be shown that  $\calt (0,1)$ can be identified with ${\rm Diff} (S^1)/SL(2,\RR)$ \cite{Verlinde:1989hv,Nag:1990dj}. The Weyl-Petersson symplectic structure reduces to the  Kirillov-Kostant \cite{Kirillov} structure on  the vacuum Virasoro coadjoint orbit, and it's quantization leads to the Virasoro vacuum  representation \cite{Witten:1987ty}.

\subsubsection*{Residual gauge fixing:  Fefferman-Graham gauge}
The residual gauge freedom  (\ref{resgauge}) can be used to bring the spatial connections $\hat A, \tilde{\hat A}$ in a desired form on the $t=0$ initial slice which we will denote as $\S_0$. Before discussing the conformal gauge, which will be used in the rest of the paper, it is instructive to first see how the standard `Fefferman-Graham gauge'   for asymptotically AdS spacetimes fits in our parametrization. This gauge leads to the (2+1)D metric in the  Banados form of \cite{Banados:1998gg}, which is an all-order version of  the  Fefferman-Graham expansion.

The Fefferman-Graham gauge corresponds to taking the 2D vielbeins to be of the form
\be 
e = {d z \over 2 y} - {y \over 2} T(x) dx, \qquad
\tilde e = {d z \over 2 y} - {y \over 2} \tilde T(x) dx.
\ee
Here, $z = x+ i y$ takes values on the upper half plane, and $T(x)$ and $\tilde T (x)$ are arbitrary functions. The corresponding spin connection one-forms are
\be 
\o = - \left( {1 \over y} + y T(x) \right) d x, \qquad \o = - \left( {1 \over y} + y \tilde T(x) \right) d x.
\ee

The vector fields $V$ and $\tilde V$ which determine the gauge choice (\ref{covgauge}) are taken to be
\be 
V = - \tilde V = - \pa_z - \pa_{\bar z} = - \pa_x.
\ee
The resulting time evolution equation (\ref{dyneqse}) is simply solved by replacing $ x \to x_+ = x +t$ in $e$ and $ x \to x_- = x -t$ in $\tilde e$.
One checks that the (2+1)D metric (\ref{3Dmetrgen}) is indeed in of Banados form  \cite{Banados:1998gg}, i.e.
\be 
ds^2 = {dy^2 + dx_+ dx_- \over y^2} - T(x_+) dx_+^2 - \tilde T(x_-) dx_-^2 + y^2 T(x_+) \tilde T(x_-) dx_+ dx_-.
\ee
The functions $- k T(x_+)$ and $- k \tilde T(x_-)$ are identified with the components of boundary stress tensor \cite{Henningson:1998gx,Balasubramanian:1999re}. 

\subsubsection*{Residual gauge fixing: conformal gauge}
Reverting to the general case where $\S_0$ has arbitrary topology,  we will  in this paper use the residual gauge freedom (\ref{resgauge}) to bring the auxiliary metrics (\ref{auxmetrs}) in the conformal gauge.
Let us  first introduce the following explicit parametrization of the zweibeins:
\be 
e = e^{- (\F+ i \l)} ( dz + \m d \bar z), \qquad \tilde  e = e^{- (\tilde  \F+ i \tilde \l)} ( dz + \tilde \m d \bar z), \label{vielb}
\ee
where $\F, \l$ are real fields and $\m$ is complex (and similarly for their tilded counterparts). We will also restrict to
\be 
|\m | <1 , \qquad |\tilde \m | <1 \label{mudisk}
\ee
so that the zweibein $(e,\bar e)$ is   invertible. 
We should keep in mind that the above is a parametrization for three-dimensional gauge potentials and that the fields $\F,\l, \m, \o $ (and their tilded counterparts) depend on all three coordinates $(t, z, \bar z)$. 

Let us now display the equations of motion in this parametrization, starting with the constraint equations (\ref{eom1}). These reduce to
\bea 
\o_z &=& -\pa_z \l - i (1- |\m|^2)^{-1}\left( (1+ |\m|^2)\pa_z\F+ \pa_{\bar z} \bar \m - \bar \m \left(\pa_z \m + 2 \pa_{\bar z} \F\right) \right)\\
e^{-2 \F} &=& { {\rm Im \,} \pa_{\bar z} \o_z \over 1- |\m|^2}
\eea
The residual gauge freedom (\ref{resgauge}) can be used  to bring both auxiliary metrics $d  s^2_2$ and $d\tilde s^2_2$ in conformal gauge on the $t=0$  $\S_0$, i.e.
\be 
\m   = \l =0, \qquad \tilde  \m   = \tilde \l  =0, \qquad {\rm at\ } t=0 \label{resgauget0}
\ee
The constraint equations (\ref{eom1}) at  $t=0$  reduce to
$\o_z  =- i \pa_z \F,\ \tilde \o_z  =- i \pa_z \tilde \F$ and Liouville's equation for $\F$ and $\tilde \F$:
\be 
 \pa_z \pa_{\bar z} \F  + e^{-2 \F}  =0, \qquad  \pa_z \pa_{\bar z}\tilde  \F  + e^{-2 \tilde \F} =0, \qquad {\rm at\ } t=0.\label{Liouveqs}
\ee
On $\S_0$, the spatial gauge field $\hat A$ takes the form
\be 
\hat A = \hat A [\F]:=  e^{-\F} dz  L_{1} -   e^{-\F} d\bar z L_{-1} + (\pa_z \F dz - \pa_{\bar z} \F d\bar z) L_0, \qquad {\rm at\ } t=0.\label{LaxA}
\ee
In this we recognize  the standard form for the Lax connection for the Liouville equation (see e.g. \cite{BabylonTalon}).
When particle sources are coupled to gravity, (\ref{Liouveqs}) will be modified by delta-function sources as we will discuss in detail in Section \ref{Secsources}. 
When the spacetime has  a conformal  boundary, these   should be supplemented with appropriate AdS boundary conditions which we will discuss in Section \ref{Secasads} below.  

From the Liouville solutions (\ref{Leq}) we can construct two holomorphic stress tensors  on the initial value surface as follows:
\be 
T(z) = -  ( \pa_z \F^2 + \pa_z^2 \F )\qquad 
\tilde T( z) = - (\pa_{ z} \tilde \F^2 + \pa_{ z}^2 \tilde \F  ) \qquad {\rm at\ } t=0. \label{TLiouv}
\ee
These will   play an important role in the rest of this work. 

Now let us consider the dynamical equations (\ref{dyneqse}) which determine the time dependence of the fields. They lead to the coupled first order equations
\bea 
\pa_t (\F + i \l) &=&   -  \left((V^z\pa_z + V^{\bar z}\pa_{\bar z})( \F + i \l)+  V^{\bar z} \pa_z \m \right)+\pa_z  {( V^z + \m  V^{\bar z}) } \nonu
\pa_t \m &=&  - \left(\pa_{\bar z} - \m \pa_ z + \pa_z \m\right) { (V^z + \m  V^{\bar z} )}.\label{dyneqs}
\eea 
These should be solved with (\ref{resgauget0}) and the solutions to (\ref{Liouveqs}) as initial conditions.
We note from (\ref{dyneqs}) that in general the fields $\m$ and $\l$ will not stay zero at later times.

\subsubsection*{Initial data for 3D gravity}

The constraint- and dynamical equations (\ref{eom1},\ref{eom2}) guarantee that the (2+1)D metric  (\ref{3Dmetrgen}) satisfies Einsteins equations with negative cosmological constant,
\be 
R_{\m\n} = - 2 g_{\m\n}.
\ee
In our parametrization (\ref{gaugepot2D},\ref{vielb}) and gauge choice (\ref{resgauget0}), the induced metric on the initial value surface $\S_0$ takes the simple form
\be 
ds^2_{\S_0} =\left( e^{- \F}  +  e^{-  \tilde  \F}  \right)^2 dz d\bar z -  \left[ \Im m \left( \pa_z (\F -\tilde \F) dz \right)\right]^2 \label{metricinit}
\ee
We note that this metric is   form-invariant  under  conformal transformations off $\S_0$ with  the Liouville fields transforming in the standard way,
\be
z \to  z'= f(z),\qquad
\F \to \F'= \F + \half \ln (f'\bar f'),\qquad
\tilde \F \to \tilde \F'=\tilde \F + \half \ln (f'\bar f').
\ee

The Liouville  (\ref{Liouveqs}) and time evolution equations (\ref{dyneqs}) ensure that the  2D metric (\ref{metricinit}) and the extrinsic curvature $K$ (whose form is rather complicated) on the initial slice satisfy the initial value constraints of (2+1)D gravity:
\bea
R^{(2)} + (K^\m_\m)^2 - K_{\m\n}K^{\m\n}+ 2 &=& 0 \nonu
D_\m K^\m_\n - D_\n K^\m_\m &=&0, \label{initGR}
\eea
where $D_\m$ is the covariant derivative with respect to the metric (\ref{metricinit}) on $\S_0$. In classical general relativity, the phase space consists of initial data satisfying (\ref{initGR}). 
%A solution to the  initial value constraints defines a  state in classical general relativity. 
In our formalism this data is repackaged  in the form of   two fields solving  Liouville's equation.  As we shall presently see, this becomes especially advantageous when including point particle matter sources. The  nontrivial problem of finding consistent initial data then reduces to solving the Liouville equation with delta-function sources, for which there exists a well-developed mathematical physics machinery.

\subsection{Spinning  particle sources   and backreaction}\label{Secsources}

We now consider gravity coupled to massive, spinning matter fields in a limit in which these fields behave like heavy point particles backreacting on the geometry. 
Let us say a little more about this point particle limit. A quantum field of mass $m$  and helicity $s$ propagates a lowest weight representation of $sl(2,\RR) \times sl(2,\RR)$ built on a primary of weight $(h, \tilde h)$, where (see e.g. \cite{Rahman:2015pzl})
\be 
m^2 = (h + \tilde h) (h+ \tilde h -2), \qquad s = h - \tilde h\label{mshhb}
\ee 
High-energy quanta are expected to behave as point particles. Since the AdS energy is given by $L_0 + \tilde L_0$, the point particle limit is  $h + \tilde h \gg 2$ and the relations (\ref{mshhb}) simplify    to
\be 
m = h + \tilde h, \qquad s = h- \tilde h.
\ee

The description of point particle sources in Chern-Simons variables goes back to \cite{Witten:1989sx}, see \cite{Ammon:2013hba},\cite{Castro:2014tta} for more recent discussions.  They are described   by a classical source action which, upon quantization, leads to a Wilson line in the appropriate unitary irreducible representation of the gauge group. A point particle  
with mass $m$ and helicity $s$ is described by the following worldline action
 \bea
 S_p [U, P,\l; A; h] &=& \int_C d\t \left[ \tr ( P D_\t U U^{-1} ) + \l \left(\tr P^2 + 2{h^2 }\right)\right]\label{Ssource}\\
 D_\t U &\equiv & {d U \over d\t} + A_\t U,\label{Spart}
 \eea
 where $A_\t$ is defined as the component of the connection parallel to $C$:
 \be
  A_\t := A_\m  (x(\t)) {d x^\m \over d\t}
  \ee
A similar action describes  the worldline coupling to $\tilde A$. 
 Here, $\t$ parametrizes the worldline $C$ and the dynamical variables $U$ and $P$ are  elements of the $SL(2,\RR)$ group and $sl(2,\RR)$ Lie algebra  respectively. % and $ c_2$ is the quadratic Casimir of the particle representation.
 The action is invariant under worldline reparametrizations, and the combined Chern-Simons and source actions remain invariant under $SL(2,\RR)$ gauge transformations $A \to \L^{-1} ( A + d) \L $ provided that the worldline fields transform as
 \be 
 U \to \L^{-1} U,\qquad  P \to \L^{-1} P \L .
 \ee
 The gauge invariance of the total action  means in particular that it is
 independent of the  shape of the curve $C$; it is therefore not necessary to vary this action with respect to the worldline $C$. 
 
 The equations of motion  following from varying the particle action with respect to the worldline variables $ U, P, \l$ reduce to
 \begin{align}
 {d P \over d\t} +[ A_\t, P]  &= 0, & {d \tilde P \over d\t} +[ \tilde A_\t, \tilde P]  &= 0 \label{CSmat1}\\
  \tr P^2 &= -2 {h^2 }, &  \tr \tilde P^2 &= -2{\tilde h^2 } \label{CSmat2} \\
 {d U \over d\t} U^{-1} +  A_\t + 2 \l P   &= 0, &  {d\tilde  U \over d\t}\tilde  U^{-1} + \tilde  A_\t + 2\tilde  \l \tilde P   &= 0\label{Ueq}
 \end{align}
 The equations for  the momenta $P$ and $\tilde P$ in the first line  
 can be shown \cite{Castro:2014tta} to be equivalent to the Mathisson-Papapetrou-Dixon (MPD) equations \cite{Mathisson:1937zz,Papapetrou:1951pa,Dixon:1970zza} governing the motion of spinning point particles in general relativity, as we review in Appendix \ref{AppMPD}.
 In the metric formulation these express conservation of momentum and angular momentum along the wordline, and generalize  the geodesic motion required for consistent coupling of non-spinning particles to gravity.
 % In particular, it is not necessary to also vary the action (\ref{Ssource}) with respect to the worldline location, as this would lead to redundant equations.
  As  explained in Appendix \ref{AppMPD}, the (2+1)D MPD equations admit solutions describing standard geodesic motion, but also allow for more general types of motion in the spinning case.  The subclass of geodesic solutions corresponds to (see  (\ref{geodsubclassPApp}))
 \be 
 [P, \tilde P] =  [E_\t, P] = [E_\t, \tilde P] =0. \label{geodsubclass}
 \ee

 Varying the total action with respect to $A$ yields the following source term for the field strength
 \be
   F_{\m\n} =  -{ \p \over k}\e_{\m\n\r} \int_C d\t   P {d x^\r \over d\t} \d^{(3)} (x - x(\t))\label{sourceq}
   \ee
 We see that, for a particle to backreact on the geometry, either $h$ or $\tilde h$ should grow linearly with $k$ in the semiclassical limit of large $k$. 
 For later convenience we define the ratios
\be 
 %\d_i = {4 \p h_ i\over k} .
 \a = {h\over k}, \qquad \tilde \a = { \tilde h\over k} .
 \ee
 It will also be useful in what follows to distinguish some special cases.
 We will call the particle `chiral' if $\a$ stays finite but $\tilde \a \to 0 $ in the large $k$ limit, and `antichiral' if the converse is true. The particle is of `generic' type if $\a\tilde \a \neq 0$.
 We will see in Section \ref{Secexs} that the case where all the particles involved are either of the chiral or anti-chiral type (i.e. $\a \tilde \a=0$) is special in that  the dynamics simplifies significantly.
 
 Let us discuss the effect of the source terms in (\ref{sourceq}) and in the analogous equation for $\tilde F_{\m\n}$ in more detail. We will restrict attention to curves $C$ for which $t (\t )$ is a monotonic function,
 \be 
 {dt \over d\t} > 0.\label{tmon}
 \ee
 We can then choose the  parameter of the curve to coincide with the time coordinate $t$,
 \be 
 x^\m (\t) = ( \t, z(\t), \bar z(\t)).
 \ee
 On general grounds, we expect that the introduction of the sources should modify the constraint equation (and hence the phase space), but not the dynamical equations.  For a generic source with $\a\tilde \a \neq 0$ this is compatible with our gauge choice (\ref{covgauge})  
 if we choose the the vectors $V$ and $\tilde V$ to coincide, on the particle worldline, with the spatial velocity
 \be 
 V ( x(t)) =  \tilde V ( x(t)) = \dot z \pa_z + \dot {\bar z}  \pa_{\bar z}, \qquad {\rm generic \ } (\a \tilde \a \neq 0).\label{Vwl1}
 \ee
 Indeed, it is straightforward to see that  the source term in (\ref{sourceq}) then does not modify the dynamic equation (\ref{eom2}). For an (anti-)chiral particle we need only impose that $V$ (resp. $\tilde V$) become equal to the spatial velocity:
 \bea
  V ( x(t)) &=&   \dot z \pa_z + \dot {\bar z}  \pa_{\bar z}, \qquad {\rm chiral \ } (\tilde \a = 0)\nonu
  \tilde V ( x(t)) &=&   \dot z \pa_z + \dot {\bar z}  \pa_{\bar z}, \qquad {\rm antichiral \ } (\a = 0).\label{Vwl2}
  \eea
 
  Note that these imply the vanishing of components of the gauge fields parallel to the worldline:
 \begin{align}
 A_\t &=  \tilde A_\t =0  &  &{\rm generic \ } (\a \tilde \a \neq 0)\nonu
  A_\t &=  0  \qquad && {\rm chiral \ } (\tilde \a = 0)\nonu
 \tilde A_\t &=  0  &&  {\rm antichiral \ }  (\a = 0).\label{Atau0}
  \end{align}
Consequently the equations (\ref{CSmat1})  impose that $P$ and $ \tilde P$ are constant. The equation (\ref{CSmat2}) can be solved as\footnote{The factors $i$ are a consequence of writing the equations in the $su(1,1)$ basis, see Footnote \ref{FNsu11}.}
  \be 
 P =- 2 h i L_0, \qquad \tilde P = - 2 i \tilde  h L_0.\label{solP}
 \ee
 The remaining  equation (\ref{Ueq}) is solved  by e.g. taking 
 \be \l=\tilde \l=0, \qquad  U=\tilde U =1.\label{sollU}\ee
 Substituting (\ref{solP}) into (\ref{sourceq}) we find that the constraint equations (\ref{Liouveqs})  on the $t=0$ slice are modified to\footnote{In our conventions, $\e_{t z \bar z}=i $ and the (2+1)D delta function is normalized such that
$1 = \int dt d^2 z \d^{(3)} (x) = 2 \int dt dx dy \d^{(3)} (x)$.}
\be \pa_z \pa_{\bar z} \F  + e^{-2 \F} = \p \sum_{i=1}^n \a_i \d^{(2)}(z - z_i), \qquad \pa_z \pa_{\bar z} \tilde \F  + e^{-2 \F} = \p \sum_{i=1}^n \tilde \a_i \d^{(2)}(z - z_i).\label{Liouveqssource}% \qquad {\rm at \ } t=0
 \ee
Here, we have performed the straightforward generalization to include $n$ particle sources labelled by $i $.   Each delta-function source term creates a deficit angle  of $2\p \a_i$ at the point $z_i$ in the 2D metric $ds^2$. We will restrict to $0 \leq \a_i  <1$ as appropriate for particle sources below the BTZ black hole treshold which corresponds to $\a=1$.

 It may seem surprising that the wordlines $C_i$ were largely arbitrary (except for the assumption (\ref{tmon})), and yet they end up satisfying the MPD equations.  The reason for this is that the vectors $V$ and $\tilde V$, which depend on the $C_i$ through (\ref{Vwl1},\ref{Vwl2}), enter in the (2+1)D metric (\ref{3Dmetrgen}) so as to ensure that the MPD equations are obeyed. We are working in a `Kantian' formulation where the object (the metric) directs itself towards our knowledge (of the worldlines) and not vice versa.
 
It is enlightening to check whether the particles in the backreacted solutions move on geodesics. Recall that this requires (\ref{geodsubclass}) to hold. In the non-chiral case, this is so due to the fact that (\ref{Atau0}) implies that the (2+1)D vielbein degenerates on the wordline:
\be 
E_\t =\half ( A_\t - \tilde A_\t)    =0 \qquad {\rm (generic)}.
\ee
The coordinate singularity signalled by the degeneration of the vielbein here reflects the conical curvature singularity on the worldline \cite{Deser:1983tn,Deser:1983nh}. 
For chiral particles however, the vielbein need not be degenerate on the worldline as 
$E_\t = \tilde A_\t/2$. They follow geodesics  if, at $t=0$,
\be 
\tilde A_\t \sim L_0.\label{chiralgeod}
\ee
If not, the chiral particle follows a more general trajectory which solves the MPD equations.

As the above analysis indicates,  the  phase space of (2+1) gravity in the presence of particle sources, is (locally) the product of  Teichm\"uller spaces of punctured Riemann surfaces,
\be 
\calt (g,b,n) \times \calt (g,b,\tilde n),
\ee
where $n$ ($\tilde n$) is the number of particles with nonvanishing quantum number $\a$ (resp. $\tilde \a$).
 For example, for $g=0,b=1,n=1$, the Teichm\"uller space $\calt (0,1,1)$ can be identified, as a symplectic manifold, with the Virasoro coadjoint orbit ${\rm Diff} S^1/U(1)$ \cite{Verlinde:1989hv},\cite{Nag:1990dj}.  Upon quantization, one obtains a generic primary representation of the Virasoro algebra. It would be of interest to have a better  mathematical understanding of the Teichm\"uller spaces with $n>1$ and $b\geq 1$.

\section{Asymptotically Anti-de Sitter spacetimes}\label{Secasads}

So far we did not specify the topology of the spatial slice $\S$.
 In this and the following   section, we will focus on the case where $\S$ has a single asymptotic boundary, which is the standard setting for the AdS/CFT correspondence. In particular we  will be describing   multi-particle excitations on a global Anti-de Sitter background.  We will comment on the case where  $\S$ has spherical topology in Section \ref{SecS2}.

When there is a asymptotic boundary, the initial value surface $\S_0$   has the topology of an open disk, possibly punctured by particle sources. The Liouville fields $\F$ and $\tilde \F$ describe hyperbolic metrics on this surface. 
We also want to  impose asymptotically AdS  boundary conditions in the standard sense of Brown and Henneaux \cite{Brown:1986nw}.
We will see that this reduces, in our parametrization,  to pseudosphere or Zamolodchikov-Zamolodchikov (ZZ) boundary conditions \cite{Zamolodchikov:2001ah} on the Liouville fields $\F$ and $\tilde \F$. Using an appropriate doubling trick we will extend all quantities to the Riemann sphere, and in particular find a connection to spherical conformal blocks. 

\subsection{AdS asymptotics and the pseudosphere}
We can model $\S_0$ as the complex upper half plane parametrized by a  complex coordinate $z$ with
\be 
{\rm Im}\, z \geq 0.
\ee
The AdS conformal boundary   corresponds to the real line, compactified  by adding the point at infinity.
 As we will see, this coordinate system will describe physics in the Poincar\'e patch. This will be  sufficient for most purposes and leads to simple formulas. The extension to a global coordinate system will be discussed at the end of this subsection. 

 In the $(t,z,\bar z)$ coordinate system the standard  boundary conditions \cite{Banados:1998gg} on the Chern-Simons gauge potentials read
\bea 
 A_- &:=& A_t -A_z- A_{\bar z}=0 \\
 \tilde A_+ &:=&  A_t +A_z+ A_{\bar z}=0 \qquad \qquad {\rm at \  Im}\, z = 0  ,\label{bc}
\eea 
In addition, we will  impose  suitable fall-off conditions \cite{Banados:1998gg}   to ensure that the resulting spacetime is asymptotically AdS.

The boundary conditions (\ref{bc}) are consistent with our gauge choice (\ref{covgauge}) provided  that the vector fields $V, \tilde V$ behave near the boundary as:
\be
V \to  - \pa_z - \pa_{\bar z}, \qquad \tilde V \to  \pa_z + \pa_{\bar z} \qquad  {\rm as\   Im}\, z \to 0.
\ee

As described in the previous section, in the presence of particle sources, we also require the vector fields $V, \tilde V$ to coincide, on the particle worldlines in the interior, with their spatial velocities (\ref{Vwl1},\ref{Vwl2}). 
Let us however first discuss the situation without sources. We can then simply take \be V = - \tilde V = - \pa_z - \pa_{\bar z}\label{VVt}\ee
throughout  the spacetime.
The constraint- and dynamical equations (\ref{Liouveqs}, \ref{dyneqs}) then reduce to
\begin{align}
0 =&\pa_z \pa_{\bar z} \F + e^{-2 \F} ,& 0=& \pa_z \pa_{\bar z} \tilde \F + e^{-2 \tilde \F} ,\label{Leq}\\
\pa_t \F =& (\pa_z +\pa_{\bar z} )\F, & \pa_t\tilde \F =& -(\pa_z +\pa_{\bar z} )\tilde \F\label{fulleqs}\\
\m =& \l = 0 , & \tilde \m =& \tilde \l =0
\end{align}
In other words, the auxiliary 2D metrics (\ref{auxmetrs}) remain in conformal gauge at all times, and the time dependence is such  that $\F$ and $\tilde \F$ depend only on the combinations $z_+ := z+t$ and $z_- := z-t$ respectively. Their real parts
$x_\pm = {\rm Re}   (z) \pm t$  will turn out to play the role of light-cone coordinates on the boundary.

The 2+1 dimensional metric (\ref{3Dmetrgen}) takes the form
\be
ds^2 
= \left| e^{- \F(z_+, \bar z_+)} d z_+   +  e^{-  \tilde  \F(z_-, \bar z_-)} d z_- \right|^2 -  \left[{\rm Im} \left( \pa_{z_+} \F(z_+, \bar z_+) d z_+ - \pa_{z_-}\tilde\F(z_-, \bar z_-) d z_-\right)\right]^2  \label{metrUHP}
\ee
One can check that the (2+1)D Einstein equations $R_{\m\n} + 2 g_{\m\n}=0$ indeed reduce to (\ref{Leq}).

An elementary solution to (\ref{fulleqs}) is to  take  $\F$ and  $\tilde \F$ to coincide at $t=0$ and take the form 
\be
\F = \tilde \F  =  \ln 2   {\rm Im}\, z,\qquad {\rm at\ } t=0.\label{AdSLiouv}
\ee
We note that $\F, \tilde \F$ diverge on the asymptotic boundary, where   $z$ is real.
The corresponding auxiliary 2D metrics (\ref{auxmetrs}) describe the hyperbolic metric on the upper half plane:
\be
 ds^2_2 =  d\tilde s^2_2 = {dz d\bar z\over 4 ({\rm Im} z)^2} . 
\ee
  The Liouville stress tensors (\ref{TLiouv}) associated to   (\ref{AdSLiouv}) vanish,
\be 
T (z) = \tilde T ( z)= 0,
\ee
and 
the (2+1)D metric (\ref{metrUHP}) is simply AdS$_3$ in Poincar\'e coordinates:
\be
ds^2 = { - dt^2 + d z d\bar z \over ({\rm Im} z)^2 }.
\ee

More generally we will consider Liouville solutions  which   approach (\ref{AdSLiouv}) near the real axis,
\be 
\F = \ln  (2 {\rm Im}\, z) + \calo (1) , \qquad \tilde \F = \ln  (2 {\rm Im}\, z) + \calo (1).\label{ZZ}
\ee
% These state that the 2D metrics approach the Poincar\'e upper half plane  metric  near the boundary. 
 In the literature on boundary Liouville theory, these are known as   Zamolodchikov-Zamolodchikov (ZZ) or pseudosphere boundary conditions \cite{Zamolodchikov:2001ah}. 
 %The solution method for this system \comment{with sources} was worked out in \cite{Hulik:2016ifr}.
 
 We will now show that the boundary conditions (\ref{ZZ})  give rise to (2+1)D metrics  obeying the Brown-Henneaux falloff conditions. As explained in  \cite{Hulik:2016ifr}, the Liouville equation determines the first subleading correction to (\ref{ZZ}) in terms of the stress tensor on the boundary:
 \bea 
 \F = \ln  (2 {\rm Im}\, z) + {2 \over 3 }  ({\rm Im}\, z)^2 T_{|z = \bar z}+ \calo ({\rm Im}\, z)^4\\
 \tilde \F = \ln  (2 {\rm Im}\, z) + {2 \over 3 } ({\rm Im}\, z)^2 \tilde T_{|z = \bar z}+ \calo ({\rm Im}\, z)^4\label{Phiexp}
\eea
The higher order terms in in this expansion are determined by  the value of the  stress tensor on the real line. The expansion (\ref{Phiexp}) and the reality of $\F, \tilde \F$ imply in particular that the value of the stress tensor on the real axis is real:
\be g
T (x) = \bar T (x), \qquad \tilde T (x) = \bar{\tilde T} (x), \qquad {\rm for\ } x\in \RR \label{Tbc}
\ee
To bring the metric in Fefferman-Graham form, we adopt following parametrization
\be 
z = x + i \left( y -{1\over 6} \left(T(x_+) + \tilde T(x_-)\right)y^3 + \calo (y^5) \right)\label{toFG}
\ee
and find
\be 
ds^2 ={dy^2 +  dx_+ dx_- \over y^2}  -  {T(x_+) } dx_+^2   - {\tilde T(x_-) } dx_-^2 + \calo (y^2) . \label{FGform} \ee
In particular, the real  functions $- kT(x_+)$ and $-k \tilde T (x_-)$ coincide with the components of the boundary stress tensor   \cite{Henningson:1998gx,Balasubramanian:1999re} of the asymptotically AdS spacetime.  We note from (\ref{toFG}) that the imaginary part of  $z$ tends to the Fefferman-Graham radial coordinate $y$ near the boundary. The  subleading correction  in (\ref{toFG} ) is 
 necessary to remove  an unwanted term of order one in $g_{yy}$. 

When particle sources are present, we can still arrange for $V$ and $\tilde V$ to take the form (\ref{VVt}) in a neighborhood of the asymptotic  boundary. Therefore the asymptotic form of the metric   (\ref{FGform}) and the relation between  the Liouville and boundary  stress tensors continues to hold in this case. Near the  delta-function sources $z_i$ in (\ref{Liouveqssource}), the Liouville field has the asymptotics
\be
\F \ \ \ \ \stackrel{\footnotesize\mathclap{\mbox{$z \to z_i$}}}{\sim}\ \ \ \  \a_j \ln |z -z_i|+\calo (1) , \qquad j = 1, \ldots, n. \label{PhinearpunctD}
\ee  
Correspondingly, the stress tensor $T(z)$ has a second order pole
\be 
T(z)  \ \ \ \ \stackrel{\footnotesize\mathclap{\mbox{$z \to z_i$}}}{\sim}\ \ \ \  {\e_j \over (z-z_i)^2} + \calo \left((z-z_i)^{-1}\right) , \qquad i = 1, \ldots, n, \label{Tnearpunct}
\ee  
where we defined 
\be 
\e_i = {\a_i \over 2} \left( 1 - {\a_i \over 2} \right).
\ee
The stress tensor $T(z)$ is therefore meromorphic  on the upper half plane. Similar properties hold  for  $\tilde \F$ and $\tilde T ( z )$.

\subsubsection*{Other useful coordinate systems}
It is often convenient to model $\S_0$   as a punctured disk instead  instead of the above  upper half plane model;  this  has the advantage that the full spacetime accessible as a bounded domain. For this purpose we make a conformal transformation on $\S_0$:
\be 
z = i {1- w\over 1- w}.\label{Cayley}
\ee
The new coordinate runs over the unit disk $|w|\leq 1$, and the AdS vacuum solution corresponds in this frame to 
\be
\F  = \tilde \F  =   \ln (1 - |w|^2),\qquad {\rm at\ } t=0.   \label{AdSdisk}
\ee
A natural choice for the vector fields $V, \tilde V$ specifying the Chern-Simons gauge choice is
\be 
V =  - i w \pa_w + i \bar w \pa_{\bar w}, \qquad  \tilde V =   i w \pa_w - i \bar w \pa_{\bar w}.
\ee
%Defining the combinations
In Appendix \ref{Appdisk} we show  that the corresponding (2+1)D metric is of the form (\ref{metrUHP}), with $z_\pm$ replaced by the combinations $w_\pm$ defined as
\be 
w_+= w e^{ i t}, \qquad w_=  w e^{- i t}.
\ee
%\be
%ds^2 
%= \left| e^{- \F(w_+, \bar w_+)} d w_+   +  e^{-  \tilde  \F(w_-, \bar w_-)} d w_- \right|^2 -  \left[{\rm Im} \left( \pa_{w_+} \F(w_+, \bar w_+) d w_+ - \pa_{w_-}\tilde\F(w_-, \bar w_-) d w_-\right)\right]^2 . \label{metrdisk}
%\ee

To conclude, we explain how to set up the equations in global coordinates. To this end we make a further conformal transformation  on $\S_0$ to a coordinate $u$ defined on the  semi-infinite cylinder,
\be 
w = e^{i u}, \qquad  {\rm Im}\, u \geq 0, \qquad   u \sim u + 2 \p .
\ee
 In the gauge determined by vector fields $V, \tilde V$ given by
\be
V =  - \pa_u - \pa_{\bar u}, \qquad \tilde V =  \pa_u + \pa_{\bar u} ,
\ee
the 2+1 dimensional metric again takes the form (\ref{metrUHP}) with $z_\pm$ replaced by $u_\pm$ defined as
%\be
%ds^2 
%= \left| e^{- \F(u_+, \bar w_+)} d w_+   +  e^{-  \tilde  \F(w_-, \bar w_-)} d w_- \right|^2 -  \left[ \Im m \left( \pa_{w_+} \F(w_+, \bar w_+) d w_+ - \pa_{w_-}\tilde\F(w_-, \bar w_-) d w_-\right)\right]^2  \label{metr}
%\ee
%where
\be 
u_+ = u + t, \qquad u_- = u-t.
\ee
Conformally mapping the AdS solution (\ref{AdSdisk}) gives
\be
\F = \tilde \F    =  \ln 2  \sinh {\rm Im}\, u\,\qquad {\rm at\ } t=0.
\ee
The stress tensors in this frame are $T (u) = \tilde T (\bar u)= {1 \over 4}$.
The (2+1)D metric is simply AdS$_3$ in global coordinates:
\be
ds^2 = d\r^2 - \cosh^2 \r dt^2 + \sinh^2 \r d\f^2.
\ee
where  we parametrized $u$ 
as
\be 
u = \f - i \ln \tanh {\r\over 2}.
\ee
% The stress tensors associated to   (\ref{AdSLiouv}) take the value
%\be 
%T (w) = \tilde T (\bar w)= {1 \over 4}.
%\ee

\subsubsection*{Pseudosphere doubling trick}

  As is usually the case for CFTs defined on a bounded domain of the complex plane with certain boundary conditions, it is possible to apply a `doubling trick' which extends the theory to the full complex plane in such a way as to automatically satisfy the boundary conditions. For the case of Liouville theory with ZZ boundary conditions this was worked out in \cite{Hulik:2016ifr}.

As we argued above, the Liouville stress tensor  $T(z)$ is   a meromorphic function on the upper half plane which, due to
the  ZZ boundary conditions,  takes on real values on the real line  (\ref{Tbc}) .  It can therefore be extended a meromorphic   function on the extended complex plane $\overline{\CC}$ satisfying
\be 
T ( z) = \bar T (z).\label{TreflUHP}
\ee
We recall the  general form   of the Liouville solution, 
\be
e^{-2 \F} = {|f'|^2 \over 4 ({\rm Im}\, f )^2}\label{Liouvsol2}
\ee 
where $f$ 
satisfies 
\be S(f, z) = 2T (z),\label{Schwarzeq}\ee 
and $S(f, z)$ denotes the Schwarzian derivative.

The reflection condition (\ref{TreflUHP}) translates into a similar reflection condition on $f$, 
\be
f(z)=  \bar  f (z).\label{reflf}
\ee
 Upon extending $f$ to the complex plane, we obtain through (\ref{Liouvsol2}) a Liouville field encoding two joined `hemispheres' of a pseudosphere, with singularities in image points, as sketched in Figure \ref{FigS2radialtime}(a).  
 \begin{figure}
%	\begin{center}
	\begin{picture}(50,200)
	\put(130,-40){\includegraphics[height=300pt]{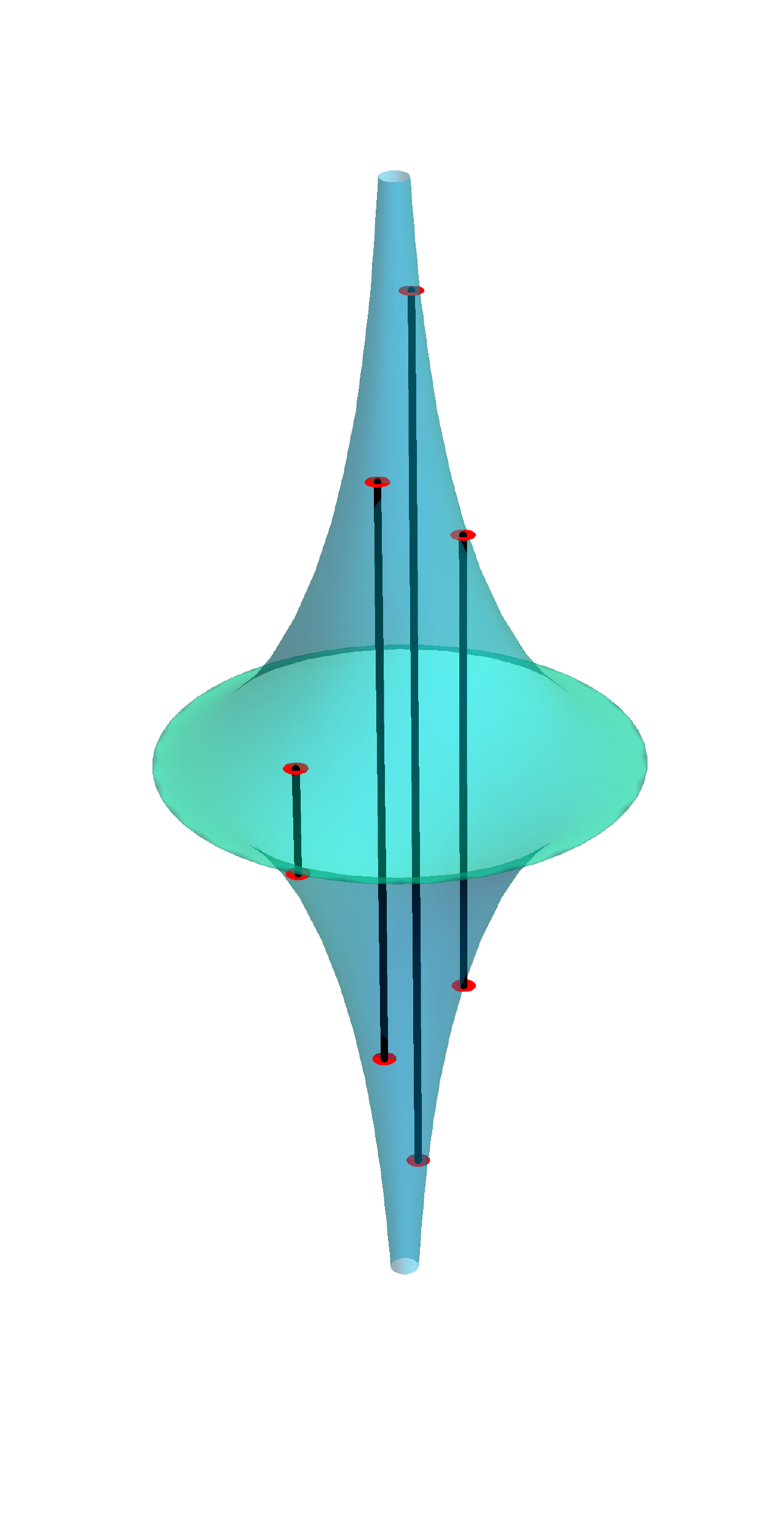}}
	\end{picture}
%		\end{center}
	\caption{The lower half of Figure \ref{FigHH}(b) can be redrawn as a filled-in pseudosphere with particle worldlines connecting image points through the bulk.}
	\label{FigpseudoS2radialtime}
\end{figure} 
In Appendix \ref{Appdisk} we give more details on the doubling trick in the unit disk conformal frame.

The function $f (z)$ in (\ref{Liouvsol2}) is multivalued  with branch points at each of the source locations. In Section \ref{Secaccess} below  we will give more details on how  $f$ is determined from the solution of a certain monodromy problem.

\subsection{Hartle-Hawking wavefunction and Liouville action}\label{SecHH}
 In this section we will consider a  Chern-Simons path integral, which is analogous to  the Hartle-Hawking \cite{Hartle:1983ai} path integral over metrics, and prepares a multi-particle state  on the disk $\S_0$ at $t=0$.  
The particles have  worldlines which  start from the conformal boundary in the past and terminate on
 $\S_0$, as in the lower half of Figure \ref{FigHH}(b).
 This  can also be depicted as in Figure \ref{FigpseudoS2radialtime}: the manifold $X$ on which we compute the Chern-Simons path integral  is a filled-in pseudosphere, where $\S_0$ is the  boundary of the northern `hemisphere'. The particle worldlines emerge from the southern hemisphere and end up at $\S_0$. We can view them as connecting image points on the pseudosphere through the bulk.  
We consider $n$ particles with quantum numbers $\a_i$ located at positions $z_i$ in the upper half plane model for $\S_0$. Their images on the southern hemisphere are located at $\bar z_i$.

 The  wavefunction of interest should be a functional of the canonical coordinate fields on  the boundary of $X$, which  are to be held fixed in the path integral. Depending on the choice of canonical coordinate one obtains   wavefunctions in different polarizations. Here we will choose a holomorphic polarization\footnote{Another common choice \cite{Verlinde:1989hv}, where $ (e, \o_z)$  are treated as coordinates and  $ (\bar e, \o_{\bar z})$  as momenta, leads to  a  different boundary term 
 $S_{bdy}= -{k \over 8 \p } \int_{\overline \CC} dz \wedge d \bar z ( |\o_z|^2 - 2 e^{-2 \F} (1- |\m|^2))$, which however coincides with (\ref{Sbdyhol}) in conformal gauge (see (\ref{Sbdy1}) below).}, where $A_z$ and $\tilde A_{\bar z}$ are treated as coordinates and   $A_{\bar z}$ and $\tilde A_{\bar z}$ as their conjugate momenta.  In order to have a consistent variational principle under the these boundary conditions, it straightforward to see that one has to add a boundary term to the action 
which can be taken to be  \cite{Moore:1989yh} 
 \be
 S_{bdy}[A]= {k \over 4 \p }\tr \int_{\overline \CC} dz \wedge d \bar z A_z A_{\bar z}.\label{Sbdyhol}
 \ee
 We note   this boundary term is actually purely imaginary. The fact that we require  a non-real boundary term is a consequence of choosing a non-real polarization.
 
The Hartle-Hawking-like path integral of interest\footnote{In principle we could have considered a more general functional which also depends on the boundary values of the worldline variable $U$, but we will specialize here to the boundary value $U=1$.} factorizes  as
\be 
\Psi_{HH} [A_z, \tilde A_{\bar z}; z_i, \bar z_i; \a_i , \tilde \a_i]
= \Psi[A_z ; z_i, \bar z_i; \a_i ]\tilde  \Psi[\tilde A_{\bar z} ; z_i, \bar z_i; \tilde \a_i ],\label{HH}
\ee
where
\be  
 \Psi[A_z ; z_i, \bar z_i; \a_i ]  =  \int_X \left. [DA DU_i DP_i D\l_i]\right|_{ A_z }e^{i \left(S_{CS}[A] + S_{bdy} [A] +\sum_i S_p[ U_i,P_i,\l_i; A;\a_i]\right)}\label{HHPsi},
%&\sim& e^{ i S_{CS}^{tot}}
% &\sim& e^{ i \left( S_{CS} [A_{cl} ] +  S_p[P_{cl}, U_{cl}, \l_{cl}; A_{cl}]\right)}.
\ee
and $\tilde \Psi$ is defined analogously. We recall that $S_p$ is the worldline action of the particles given in (\ref{Spart}).   

In the standard approach to quantizing Chern-Simons theory with compact gauge groups on compact manifolds, a similar functional gives a path integral representation of Kac-Moody blocks \cite{Labastida:1989wt}.   For noncompact gauge groups and the extension to  manifolds $\S_0$ with asymptotic boundary, the exact evaluation of path integrals of the form (\ref{HHPsi}) is progressively less straightforward,   see \cite{Porrati:2019knx} for  a recent discussion. In this work, we will focus on a more pedestrian goal and consider the large $k$ limit where  (\ref{HHPsi}) is dominated  by a classical saddle point, which we will then relate to the large $k$ limit of an appropriate CFT block.
At large $k$, the wavefunction $\Psi [A_z; z_i, \bar z_i; h_i ]$ is well approximated by the classical action 
\be  
 \Psi[A_z; z_i, \bar z_i; \a_i ] \ \ \ \ \stackrel{\footnotesize\mathclap{\mbox{$k \to \infty$}}}{\sim}\ \ \ \  e^{ i S_{cl}^{tot}} ,
 \ee
where the total action $S^{tot} =  S_{CS}  + S_{bdy} + \sum_i S_{p,i}$ includes  the boundary and  source terms.
It is to be evaluated on the classical solution
which takes on the specified value $A_z$ on the boundary.
 
 We will first show that the total on-shell action receives contributions only from the boundary term $S_{bdy} $ in (\ref{Sbdyhol}).
 Recalling that, in our chosen gauge, the component $A_\t$ parallel to the worldline vanishes, and using the solution  (\ref{solP},\ref{sollU}) for the worldline fields, one sees that the source action $S_p$ vanishes.
 The bulk Chern-Simons action can be  rewritten  as
\be 
S_{CS}[A] = {k \over 4\p} \tr \int_X \left( A\wedge F - {1\over 3} A \wedge A \wedge A  \right).
\ee
The second term vanishes since $\det A^a_\m=0$ in our gauge (\ref{covgauge}), while the first term vanishes upon  using (\ref{sourceq}) and (\ref{Atau0}).
Therefore only the boundary term  contributes to the total action and gives
\be 
 \Psi[A_z; z_i, \bar z_i; \a_i ] \sim \exp {ik \over 4 \p }\tr \int_{\overline \CC} dz \wedge d \bar z A_z A_{\bar z}^{cl}.
 \ee
Here, $A_{\bar z}^{cl}$ is a solution to
\be 
\pa_z A_{\bar z}^{cl} -\pa_{\bar z} A_z + [A_z, A_{\bar z}^{cl}] = - 2\p \sum_i \a_i \d^{(2)} (z-z_i)  L_0.\label{constrHH}
\ee
 From this expression, we learn that the argument $A_z$ is cannot completely arbitrary:  the wavefunction has support on those $A_z$ which arise as components of an almost everywhere flat connection\footnote{This is a  well-known fact \cite{Elitzur:1989nr} which, from the form (\ref{Scandec}), can be seen to extend to the full quantum functional integral. }.   In other words, $A_z$ must locally be of the form $A_z = G^{-1} \pa_z G$, and due to (\ref{constrHH}) we must have, locally and in the $sl(2,\RR)$ basis,
 $ A_{\bar z}^{cl}=G^{-1} \pa_{\bar z} G$. In order to match up with our conformal gauge choice on $\S_0$ we  we will further specialize the wavefunction to those $A_z$ which are of the  form (see (\ref{LaxA}))
\be 
A_z =   e^{-\F}   L_{1}  + \pa_z \F  L_0 \Rightarrow A_{\bar z}^{cl} =  - e^{-\F}   L_{-1}  - \pa_{\bar z} \F  L_0 ,\label{special}
\ee
where $\F$ satisfies (\ref{Liouveqssource}). The resulting specialized wavefunction then depends only on $z_i, \bar z_i; \a_i $ and we will write it as
$\Psi [ z_i, \bar z_i; \a_i ]$ in what follows.
 
 Summarizing, we have argued that  
  \be
\Psi[z_i, \bar z_i; \a_i ]\sim e^{ i S_{bdy}'[\F]},  
 \ee
 where
  \be
 S_{bdy}'[\F]=  -{ k\over 8\p} \int_{\overline{\CC}} dz \wedge d \bar z \left( \pa_z \F \pa_{\bar z} \F
 - 2 e^{-2 \F} \right)\label{Sbdy1}
 \ee
 Our final result   will contain two corrections to this expression. First of all, (\ref{Sbdy1}) is not finite due to divergent contributions from  the punctures. To regularize it, we will replace the integration region by $\overline{\CC}_\e$, which is the extended complex plane $\bar \CC$  with small discs of radius $\e$ removed around each of the source insertions $z_j$ and their images $\bar z_j$, as well as
  small strips of width $\e$ above and below the real axis. In order to render the action finite upon taking the limit $\e \to 0$, we see from the  behavior (\ref{PhinearpunctD}) near the punctures that we should add the following term to (\ref{Sbdy1}): 
 \be 
 r_\e = - 2 \p   \sum_{j=1}^{n} \a_j^2   \ln \e.\label{repsUHP}
 \ee
 Note that it is independent of the source locations $z_i$.
 
 A second modification is due to the fact that  (\ref{Sbdy1}) is not differentiable as a functional of $\F$, again  due to the source terms. The need to insist on differentiability was stressed in a related context in \cite{Brown:1986ed}.
 The variation of the integral (\ref{Sbdy1})  picks up  boundary terms in the domain $\overline{\CC}_\e$
 \be 
 \d S_{bdy}'[\F] =  -{ k\over 8\p} \lim_{\e \to 0}\int_{\d \overline{\CC}_\e}  \d\F \left(
\pa_{\bar z} \F d\bar z - \pa_z \F d z\right)+\ldots\label{singvar}
%\d S_{bdy}'[\F] =  -{ k\over 8\p} \lim_{\e \to 0}\sum_{j=1}^n\left(  \oint_{C_j^\e} +  \oint_{\tilde C_j^\e} + \right) \d\F \left(
%\pa_{\bar z} \F d\bar z - \pa_z \F d z\right)\label{singvar}
  \ee 
  This variation can be cancelled on fields which behave as (\ref{ZZ},\ref{PhinearpunctD}) by adding  contour integrals  along the boundary components  of $\overline{\CC}_\e$.
   Upon doing so we arrive at the following finite and differentiable boundary action    
      \bea 
 S_{bdy}^\e[\F] 
  &=&  {k \over 8 \p}  \left(  \int_{\overline{\CC}_\e} dz \wedge d \bar z \left( |\pa_z \F|^2- 2   e^{-2 \F} \right)+ r_\e\right. \\
 && - \sum_{j=1}^{n}{\a_j \over 2} \left(\oint_{C_j^\e} \F \left( {d\bar z\over \bar z - \bar z_j} - {d z\over  z -  z_j} \right) +  \oint_{\tilde C_j^\e} \F \left( {d\bar z\over \bar z -  z_j} - {d z\over  z - \bar z_j} \right) \right)\\
&&\left. + \left(  \int_{\RR + i \e} + \int_{\RR - i \e} \right) \F   {d \bar z + dz  \over z - \bar z} \right)\label{Sbdyeps}
\eea 
  Here, the $C_j$  and $\tilde C_j$ are  circular  contours of radius $\e$ around the source in $z_j$ and its image in $\bar z_j$ respectively. All line integrals in the above expression are  oriented as boundary components of $\overline{\CC}_\e$.
  
 Our main observation is now that the functional (\ref{Sbdyeps}) closely resembles the
 standard, regularized, Liouville action  on the pseudosphere $S^\e_L [\F]$, see \cite{Zamolodchikov:2001ah}, 
 except for a wrong coefficient multiplying  the Liouville potential $e^{- 2 \F}$. More precisely we have 
\be 
i S_{bdy}^\e[\F]  = -  {k \over 4} S_L^\e [ \F] + {3 i k\over 8 \p }  \int_{\overline{\CC}_\e}  dz \wedge d \bar z  e^{-2 \F} .
 \ee
Using the fact the source-free Liouville equation holds on $\overline{\CC}_\e$,  making  use of (\ref{ZZ},\ref{PhinearpunctD})   and taking the $\e \to 0$ limit, we find that
\be 
i S_{bdy}[\F] =-  {k \over 4} S_L[ \F_{cl}] - {3  k\over 4  } \sum_{j=1}^n \a_j  .\label{CSLiouvUHP}
 \ee
 Therefore we have established that the on-shell Chern-Simons action on $X$ is  proportional to the Liouville action on the pseudosphere, up to a term independent of the insertion points $z_i$. 
We conclude that the classical approximation to the Hartle-Hawking wavefunction (\ref{HH}) is of the form 
  \be 
 \Psi ( z_i, \bar z_i, \a_i ) \sim N (\a_i, \tilde \a_i ) e^{ - {k \over 4} (S_L [\F_{cl} (z_i,\bar z_i, \a_i)] + S_L [\tilde \F_{cl} (z_i,\bar z_i,\tilde \a_i)] )}.\label{HHSL}
 \ee
 with $N(\a_i, \tilde \a_i)$ independent of the insertion points $z_i$.

\subsection{Accessory parameters and pseudosphere Polyakov relations}\label{Secaccess}
The property (\ref{HHSL}) will   allow us to relate the wavefunction to a classical Virasoro block. An important ingredient for this connection  is a  property of the on-shell Liouville action known as the Polyakov relation.  We start by  recalling,  following \cite{Menotti:2006gc,Hulik:2016ifr}, the role of accessory parameters in the monodromy problem involved in solving the sourced Liouville equation (\ref{Liouveqssource}) on the upper half plane with boundary conditions (\ref{ZZ}).
 We will then fill a gap in the literature and derive a Polyakov relation for Liouville theory on the pseudosphere, identifying the on-shell action as the generating function of the accessory parameters.
%By making an $SL(2,\RR)$ transformation  we can fix $z_n$ and the imaginary part of  $z_{n-1} \in \RR$. 
\subsubsection*{Acessory parameters}
From (\ref{PhinearpunctD}) and regularity at infinity, it follows that the stress tensor of a solution to (\ref{Liouveqssource}, \ref{ZZ}) is a meromorphic function on $\overline{\CC}$ of the form
 \be
T (z) =\sum_{i=1}^{n}  \left({\epsilon_i \over  (z-z_i)^2}+ {   \epsilon_i \over  (z-\bar z_i)^2}+ {c_i\over z-z_i}+ {\tilde c_i\over z-\bar z_i}\right)   \label{accessparD}
\ee
The as yet undetermined coefficients $c_i$ and $\tilde c_i$ are called  accessory parameters. They are constrained by the reflection condition (\ref{TreflUHP}) as well as the requirement of regularity as $z \to \infty$. This leads to $ 2n + 3$ real conditions on the accessory parameters, namely
\bea
\tilde c_i  &=& \bar c_i, \qquad i = 1, \ldots, n \label{ccond1}\\
{\rm Re} \left(  \sum_{i=1}^{n} c_i \right) &=&0\label{ccond2}\\ 
 {\rm Re} \left(  \sum_{i=1}^{n} (c_i z_i + \e_i )\right)&=&0\label{ccond3}\\
  {\rm Re} \left(  \sum_{i=1}^{n} (c_i z_i^2 + 2 \e_i z_i)\right)&=&0 \label{ccond4}
\eea
Upon imposing these we are left with a $2n-3$ real-dimensional space of undetermined  accessory parameters.

\subsubsection*{Liouville monodromy problem}
The Liouville solution is determined by a multivalued function $f(z)$ as in  (\ref{Liouvsol2}). The Schwarzian derivative of this function is proportional to the stress tensor, cfr. (\ref{Schwarzeq}). 
In practice, the function $f(z)$ is constructed from solutions to the ordinary differential equation (ODE)
\be 
(\pa_z^2 + T(z) )\psi (z) =0.\label{ODET}
\ee
Letting $\psi_1$ and $\psi_2$ denote solutions to with unit Wronskian, i.e. $\psi_1'\psi_2 - \psi_2'\psi_1 =1$, the function $f$ is the ratio
\be 
f = {\psi_1 \over \psi_2}.
\ee
We can also write (\ref{Liouvsol2}) in terms of $\psi_{1,2}$ as
\be 
e^{\F} = i (\psi_1 \bar \psi_2- \bar \psi_1 \psi_2 )\label{Phipsis}
\ee
For general accessory parameters,  the vector $(\psi_1\ \psi_2)^T$ comes back to itself  modulo an $SL(2,\CC)$ monodromy matrix when encircling a singular point, while $f$ transforms by the associated fractional linear transformation.  
However, the expression (\ref{Liouvsol2}) is only invariant under the $SL(2, \RR)$ subgroup of $ SL(2,\CC)$. Therefore, in order for $e^\F$ to be  single-valued, 
the    accessory parameters must be chosen so as to restrict all monodromies to lie in  $SL(2, \RR) \subset SL(2,\CC)$.  It can be shown \cite{Hulik:2016ifr} that this requirement  imposes $2n-3$ real conditions, precisely as many as the number of undetermined accessory parameters. Even though there is no known  proof of existence and uniqueness for the Liouville solution with pseudosphere boundary conditions (as far as we know), one therefore expects that at least for some values of the parameters  a solution can exist. Below we will derive a more detailed existence condition in the form of a reflection property of a  conformal block on the sphere.

\subsubsection*{Polyakov relation on the pseudosphere}

As it turns out, the solution of the above monodromy problem, if it exists, is fully determined by the on-shell Liouville action. In the case of Liouville theory on the sphere, this is a well-known property which follows from relations  originally conjectured by Polyakov and subsequently proven in \cite{Zograf_1988,Takhtajan:2001uj}.
We now   show that also on the pseudosphere  a Polyakov-type relation holds in the sense that the on-shell Liouville action is a generating function for the accessory parameters:
\be 
{\pa S_L \over \pa z_i }=2 c_i, \qquad {\pa S_L \over \pa \bar z_i }=2 \tilde c_i\label{SLcPS2}
\ee
The factor of 2 in these expressions can be thought of as coming from contributions from the image charges.
We include the derivation here for completeness since we are not aware of its appearance elsewhere.
We follow closely the method  used in \cite{Takhtajan:2001uj} for  Liouville theory on the sphere.  We want to compute
\be 
\lim_{\e \to 0} \pa_{z_i} S_L^\e,
\ee
where we recall (cfr. (\ref{CSLiouvUHP})) that  $S_L^\e$ is given by 
  \bea 
 S_{L}^\e[\F] 
  &=&  {i \over 2 \p}  \left(  \int_{\overline{\CC}_\e} dz \wedge d \bar z \left( |\pa_z \F|^2 - 2 e^{-2 \F} \right)+ r_\e\right. \\
 && - \sum_{j=1}^{n}{\a_j \over 2} \left(\oint_{C_j^\e} \F \left( {d\bar z\over \bar z - \bar z_j} - {d z\over  z -  z_j} \right) +  \oint_{\tilde C_j^\e} \F \left( {d\bar z\over \bar z -  z_j} - {d z\over  z - \bar z_j} \right) \right)\\
&&\left. + \left(  \int_{\RR + i \e} + \int_{\RR - i \e} \right) \F   {d \bar z + dz  \over z - \bar z} \right),
\eea 
with $r_\e$ given in (\ref{repsUHP}). In computing the derivative with respect to $z_i$ we have to take into account contributions coming from varying the integration domain, which can be  converted into derivatives of step functions  $\theta ( |z -z_i|- \e)$ in the integrand. This leads to the identity
\be 
\pa_{z_i} \int_{\overline{\CC}_\e} dz \wedge d \bar z \, G = \oint_{C_i^\e} G\, d\bar z -\oint_{\tilde C_i^\e} G\, d z  + \int_{\overline{\CC}_\e} dz \wedge d \bar z\, \pa_{z_i}  G.\label{dersdomain}
\ee
Also,  the  Liouville equation implies that, on $\overline{\CC}_\e$,
\be 
\pa_{z_i} \left( |\pa_z \F|^2 + e^{-2\F} \right) = \pa_z \left( \pa_{z_i} \F \pa_{\bar z} \F \right) +
\pa_{\bar z} \left( \pa_{z_i} \F \pa_{ z} \F \right).
\ee
Making use of these identities  we find
\bea
 \pa_{z_i} S_L^\e &=&  {i \over 2 \p} \left( \oint_{C_i^\e} \left( |\pa_z \F|^2 + e^{-2 \F} \right)  d\bar z -\oint_{\tilde C_i^\e} \left( |\pa_z \F|^2 + e^{-2 \F} \right)  d z \right. \nonu
 &&+ \sum_{j=1}^n  \left( \oint_{C_j^\e}+ \oint_{\tilde C_j^\e} \right) \pa_{z_i}\F \left(\pa_{\bar z}\F d\bar z - \pa_z\F dz\right)\nonu
 && -  \sum_{j=1}^{n} {\a_j \over 2} \oint_{C_j^\e}
 \left( \pa_{z_i} \F + \d_{ij} \pa_z \F \right) \left( {d\bar z\over \bar z - \bar z_j} - {d z\over  z -  z_j} \right)\nonu
 && -  \sum_{j=1}^{n} {\a_j \over 2} \oint_{\tilde C_j^\e}
 \left( \pa_{z_i} \F + \d_{ij} \pa_{\bar z} \F \right) \left( {d\bar z\over \bar z -  z_j} - {d z\over  z - \bar z_j} \right)\nonu
 && \left. + \left(  \int_{\RR + i \e} + \int_{\RR - i \e} \right)  \pa_{z_i} \F   {d \bar z + dz  \over z - \bar z} \right),\label{delSLD}
 \eea
 To evaluate this in the $\e \to 0$ limit  we  need the first subleading terms in the expansions near the boundary, already derived in  (\ref{Phiexp}), and near the punctures
 (\ref{PhinearpunctD}). The latter are determined by the form of the stress tensor (\ref{accessparD}) and one finds 
\bea 
\F &\ \ \ \ \stackrel{\footnotesize\mathclap{\mbox{$z \to z_j$}}}{\sim}\ \ \ \ &  \a_j \ln |z -z_j|+ \s_j - {c_j \over \a_j}(z-z_j) - {\bar c_j \over \a_j}(\bar z-\bar z_j) +\ldots, \qquad j = 1, \ldots, n \label{PhinearpunctD2}\nonu
&\ \ \ \ \stackrel{\footnotesize\mathclap{\mbox{$z \to \bar z_j$}}}{\sim}\ \ \ \ &  \a_j \ln |z -\bar z_j|+\tilde \s_j - {\bar c_j \over \a_j}(z-\bar z_j) - { c_j \over \a_j}(\bar z- z_j) +\ldots, \qquad j = 1, \ldots, n \label{PhinearpunctD3}\nonu
 &\ \ \ \ \stackrel{\footnotesize\mathclap{\mbox{${\rm Im}\, z \to 0$}}}{\sim}\ \ \ \  & \ln (2 {\rm Im}\ z) + {2 \over 3}  ( {\rm Im}\, z)^2 T_{|z = \bar z} +\ldots,\label{Phinearbdy2}
\eea
where the $\s_i$ are functions of $(z_j, \bar z_j )$.  Using these expansions in (\ref{delSLD})   and taking the $\e \to 0$ limit we obtain 
\bea 
\lim_{\e \to 0} \pa_{z_i} S_L^\e &=& - c_i\nonu
&& + \sum_{j=1}^{n} \a_j (\pa_{z_i} \s_j + \pa_{z_i} \tilde  \s_j)   + {3 } c_i\nonu
&& - \sum_{j=1}^{n} \a_j \pa_{z_i} \s_j \nonu
&&  - \sum_{j=1}^{n} \a_j \pa_{z_i} \tilde  \s_j\nonu
&& + 0 .
\eea
Each line in this expression is the contribution  of the corresponding line in (\ref{delSLD}).
%-  \sum_{j=1}^n \a_j \s_j + 2 \s_n  .
Adding these up we arrive at (\ref{SLcPS2}).

 \subsection{Wavefunctions and solutions from  vacuum blocks on the sphere}
 We are now ready to illustrate the intimate connection with classical Virasoro blocks on the sphere.
It was  argued in \cite{Hulik:2016ifr} that {\em if} a solution to  (\ref{Liouveqssource})  with boundary conditions (\ref{ZZ}) exists, it is closely related to, and determined by, a specific classical Virasoro vacuum block. We will now review and extend this argument. 
h
\subsubsection*{Monodromy problem for classical blocks}\label{blockproblem}
We start by recalling   some properties of classical Virasoro blocks and their construction through monodromy methods \cite{Zamblock}, following closely \cite{Hartman:2013mia}.
 Quantum conformal blocks $\calf (z_I, \D_I, \D_{J'})$ are basic building blocks of CFT correlators which capture the parts which are  fixed by Virasoro symmetry. They depend on the dimensions $\D_I, I = 1, \ldots, m$ of the primary operators as well as on dimensions $\D_{J'}, J' = 1, \ldots , m-3$ of the exchanged conformal families. They also depend implicitly on the chosen OPE channel in which the exchanged  families propagate.

If we take the classical $ c= 6k \to \infty$ limit with $\e_I = \D_I /k, \n_{J'} = \D_{J'}/k$ fixed, it has been argued that the conformal block exponentiates as follows
\be 
\calf (z_I, \D_I, \D_{J'}) \ \ \ \ \stackrel{\footnotesize\mathclap{\mbox{$k \to \infty$}}}{\sim} \ \ \ \  e^{ - k F ( z_I, \e_I, \n_{J'})},
\ee
where $F$ is called a classical Virasoro block. Let us denote by $D_I$ the partial derivatives
\be 
\pa_{z_I}  F ( z_I, \e_I, \n_{J'}) \equiv D_I ( z_i, \e_i, \n_{J'}).\label{confWard}
\ee
The conformal Ward identities imply that the $D_I$ play the role of accessory parameters in a meromorphic stress-energy tensor
\be 
t (z) = \sum_{I=1}^{m}\left( {\e_I \over  (z-z_I)^2}+ {D_I \over z-z_I}\right),\label{tS2}
\ee
and satisfy constraints from regularity of $t(z)$ at infinity:
\bea 
\sum_I D_I & = & 0\label{Cconstr1} \\
\sum_I (D_I z_I + \epsilon_I) & = & 0 \label{Cconstr2}\\
\sum_I (D_I z_I^2 + 2\epsilon_I z_I) & = & 0.\label{Cconstr3}
\eea
The classical block accessory parameters $D_I$ are determined \cite{Hartman:2013mia} by requiring  
that the solutions to the ordinary differential equation
\be 
( \pa_z^2 + t(z) )\psi =0
\ee
have a monodromy matrix $M_{J'}$ when going around  the $J'$-th  closed loop in the conformal block diagram, whose conjugacy class  is fixed by $\n_{J'}$:
\be 
\tr M_{J'} = -2\cos ( \p \sqrt{ 1- 4 \n_{J'} }).
\ee

\subsubsection*{The wavefunction as a classical block}
The following property  will allow us to identify the  accessory parameters $c_i$ appearing in the Liouville monodromy problem with  accessory parameters $D_I$ for a certain classical block. 
Suppose that a solution  to the inhomogeneous Liouville equation (\ref{Liouveqssource}) on the upper half plane exists  under boundary conditions (\ref{ZZ}). Then it is easy to see from the reflection property (\ref{reflf}) that the  monodromies of the multivalued function $f$ around image points are each others' inverse:
\be 
M_{(z_i, \bar z_i)} = \left( M_{(\bar z_i,  z_i)} \right)^{-1}.\label{monim}
\ee
Indeed, let  $M$ denote the monodromy matrix that $f$ picks up when encircling the point  $(z_i, \bar z_i)$ counterclockwise. The reflection property (\ref{reflf}) then implies that $\bar M$ is  the monodromy picked up when clockwise encircling the image point $(\bar z_i,  z_i)$. Since $f$ determines a Liouville solution, all monodromies must lie in $SL(2,\RR )$ so that $\bar M = M$. The monodromy when encircling the image point counterclockwise is therefore $M^{-1}$ as advertised.

\begin{figure}
	\centering
		\includegraphics[scale=0.15]{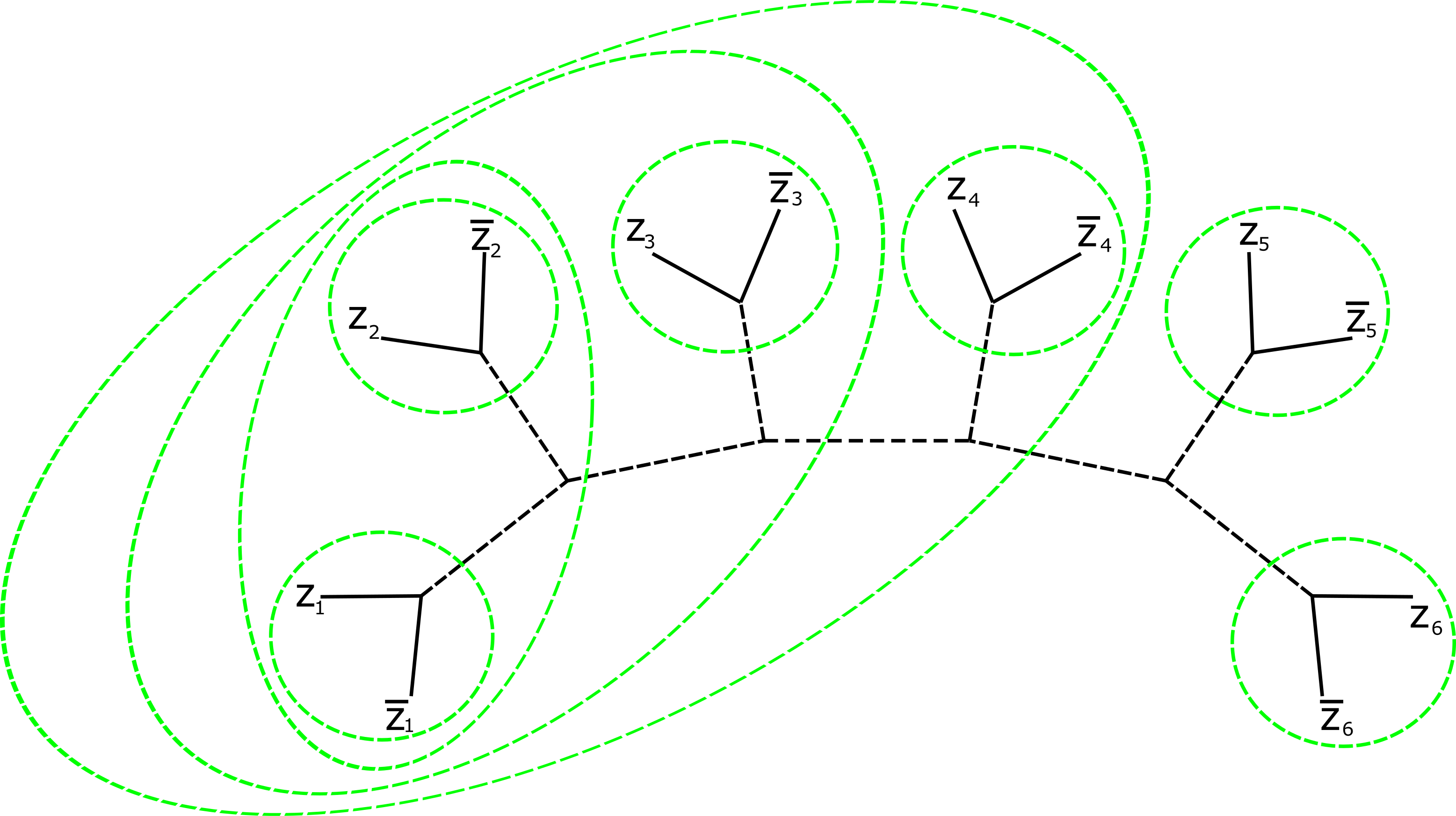}
	\caption{The  OPE channel relevant for the definition of the vacuum block $F_0$. Solid lines indicate external mirrored operators which are fused together to an identity operator. The green dashed circles indicate cycles of trivial monodromy.}
        \label{fig:OPEchannel}
\end{figure}

The property (\ref{monim}) implies that monodromy of solutions to the ODE (\ref{ODET}),
 when encircling any pair of image points, is trivial. 
From the above discussion of classical blocks  we then see   that the Liouville accessory parameters $c_i, \tilde c_i$ solve the monodromy problem determining 
 a $2n$-point  conformal block on the sphere. The primary operators are inserted pairwise in image points, and the relevant OPE  channel is the one which fuses image pairs, as illustrated in Figure \ref{fig:OPEchannel}.
The triviality of the monodromy around image points tells us that this block is a vacuum block. We will use the shorthand notation $F_0 (z_i,\bar z_i, \e_i)$ to denote this block, i.e.
 \be 
  F_0 (z_i,\bar z_i, \e_i) := F (z_i, z_{n+i}= \bar z_i; \e_i , \e_{n+i} = \e_i; \n_{J'}=0).\label{defF0}
 \ee
It generates the Liouville accessory parameters through (\ref{confWard}),
\be 
\pa_{z_i} F_0(z_i,\bar z_i, \e_i) = c_i,\qquad \pa_{\bar z_i} F_0(z_i,\bar z_i, \e_i) = \tilde c_i.\label{cderF}
\ee
Since the doubling trick for the pseudosphere  requires that $\tilde c_i = \bar c_i$  (see (\ref{ccond1})), we see that a necessary condition for the Liouville solution to exist is that the block $F_0 (z_i,\bar z_i, \e_i)$ is, up to an irrelevant constant  independent of the insertion points, a real function,
\be 
 F_0(z_i,\bar z_i, \e_i)  \in \RR.\label{existencecrit}
 \ee
 Note that the remaining conditions on the accessory parameters $c_i$ (\ref{ccond2}-\ref{ccond4}) are then implied by (\ref{Cconstr1}-\ref{Cconstr3}).

Let us analyze the existence  criterion (\ref{existencecrit}) in more detail.
 {In general the conformal block, whether semi-classical or quantum, is a function purely of the cross-ratios, scaling weights, exchange conformal weights and central charge. As is well-known the conformal block in a fixed channel can be constructed level-by-level in terms of descendent states. 
 
 In the case of the four-point function of primary operators $\langle \phi_1(0)\phi_2(x)\phi_3(1)\phi_4(\infty)\rangle$ the conformal block for an exchanged primary of holomorphic weight $\D_p$ takes the form}
\begin{equation}
\mathcal{F}^{12}_{34}(x;p) = \sum_{K=0}^{\infty}\sum_{\{k\}} \beta^{\{k\}}_{12p} x^{K} \frac{\langle [L,...[L,\phi_p(0)]...]\phi_3(1)\phi_4(\infty)\rangle}{\langle \phi_p(0)\phi_3(1)\phi_4(\infty)\rangle}.  
\end{equation}
{Here the sum over $\{k\}$ indicates a sum over all partitions of the integer $K$, the numbers $\beta^{\{k\}}_{12p}$ are the components of the linear decomposition of the OPE into descendent states. Both the numbers $\beta^{\{k\}}_{12p}$ as well as the ratio of 3pt.-functions are fixed entirely by the Virasoro algebra and furthermore they can be constructed entirely by algebraic operations on the scaling dimensions and central charge, as a result these factors are manifestly real. As a consequence, the burden for the reality condition lies on the conformal cross-ratio $x$: as long as $x$ is positive and real the conformal block will be real.} %\textbf{Gideon: Something we ignored before, because the scaling weights are not integers the cross-ratio should be positive as well, this doesn't add any additional constraints though.}
For operators inserted in image points as in (\ref{defF0}), the conformal cross-ratio takes the form
\begin{equation}
x=\frac{(z_1-\bar{z}_1)(\bar{z}_2-z_2)}{(z_1-\bar{z}_2)(\bar{z}_1-z_2)},
\end{equation}
{which is manifestly real and positive. Hence in the case of four-points the reality condition imposes no additional constraints. 

The situation changes once one considers six points. In this case the relevant cross-ratios are given by}
\begin{equation} 
z=\frac{(z_5-z_6)(z_1-z_4)}{(z_4-z_5)(z_6-z_1)}, \;\; u=\frac{(z_5-z_6)(z_2-z_4)}{(z_4-z_5)(z_6-z_2)}, \;\; v=\frac{(z_5-z_6)(z_3-z_4)}{(z_4-z_5)(z_6-z_3)}.
\end{equation}
In the OPE-channel where operators are contracted with the ones located at their mirror points, the so-called `star'-channel in the case of six points \cite{Anous:2020vtw}, the conformal block takes the form
\begin{align}
& \mathcal{F}^{12}_{34 56}(z,u,v;p,q,s)=\sum_{K_1,K_2,K_3}\sum_{\{k_1\},\{k_2\},\{k_3\}} z^{K_1}u^{K_2}v^{K_3} \beta_{12p}^{\{k_1\}}\beta_{34q}^{\{k_2\}}\beta_{56s}^{\{k_3\}} \nonumber \\
& \times\frac{\langle [L,...[L,\phi_p(0)]...][L,...[L,\phi_q(1)]...][L,...[L,\phi_s(\infty)]...]\rangle}{\langle \phi_p(0)\phi_q(1)\phi_s(\infty)\rangle}.
\end{align}
{By the same argument as before the burden for reality lies on the conformal cross-ratios, the exception now being that the reflection symmetry alone (i.e. $z_4=\bar{z}_1$, $z_5=\bar{z}_2$, $z_6=\bar{z_3}$) is not sufficient to ensure that all three cross-ratios are real and positive. In conclusion, for 3 or more particles on the pseudoshere the existence criterion (\ref{existencecrit}) can be met only for the  restricted particle  positions for which all cross-ratios are real and positive.
 
Returning to the property (\ref{cderF}), comparing it  with the Polyakov relation (\ref{SLcPS2}) we see that the Liouville action and the classical block must be proportional (up to an irrelevant additive constant):
\be 
 S_L [\F_{cl} (z_i,\bar z_i, \a_i)] = 2 F_0(z_i,\bar z_i, \e_i) .
 \ee
 Combining with a similar result for the $\tilde \F$  Liouville field and our expression (\ref{HHSL}) for the wavefunction in terms of the Liouville action, we obtain the following expression for the classical approximation to the  wavefunction (\ref{HHPsi}):  \be 
 \Psi_{HH} ( z_i, \bar z_i, \a_i ) \sim N (\a_i, \tilde \a_i ) e^{ - {k \over 2} (F_0 ( z_i, \bar z_i, \a_i ) +F_0 ( z_i, \bar z_i,\tilde  \a_i ) )}.
 \ee

Furthermore, the knowledge of the classical block $F_0$ also determines, in principle, the full (2+1)D backreacted metric:  the metric on $\S_0$ is constructed from solutions to two  ODE's of the form (\ref{ODET}) with coefficients determined by  $F_0$, and its subsequent time evolution is determined by the first order system (\ref{dyneqs}).

\subsection{Further developments}
Before discussing special cases and explicit examples, we would like to clarify two aspects of our formalism. The first is how our framework is related to the description of (2+1)D gravity in terms of a single Liouville field living on the boundary. The second is a concrete description of the improper gauge transformations which `add boundary gravitons' to the solution. 
\subsubsection*{Reconstruction of the Lorentzian boundary Liouville field}
It is well-known from the work \cite{Coussaert:1995zp} of Coussaert, Henneaux and van Driel that (2+1)D AdS gravity in the presence of a single conformal boundary can be reformulated in terms of a Lorentzian\footnote{See \cite{Krasnov:2000zq,Krasnov:2000ia} for the role of boundary Liouville theory in Euclidean spacetinmes.} Liouville field $\F_B$ living on the   boundary. In this paragraph we wish to clarify the relation between this description and our parametrization 
in terms of two Euclidean Liouville fields $\F, \tilde \F$ defined on the initial  slice $\S_0$.  

The boundary Liouville field $\F_B$ of \cite{Coussaert:1995zp} satisfies the Lorentzian equation
\be
\pa_+ \pa_- \F_B +e^{-2 \F_B} =0,\label{LiouvLor}
\ee
where $x_+, x_-$ are boundary lightcone coordinates. A solution to this equation determines an asymptotically AdS gravity solution whose boundary stress tensor components (rescaled by a factor $-(2 k)^{-1})$) are
\be 
T (x_+ ) = - \left( (\pa_+ \F_B)^2 + \pa_+^2 \F_B \right), \qquad \tilde T (x_- ) = - \left( (\pa_- \F_B)^2 + \pa_-^2 \F_B \right)\label{TB}
\ee
The general solution to (\ref{LiouvLor}) can be expressed in terms of two real  functions $f_B(x_+)$ and $\tilde f_B(x_-)$ as 
\be 
e^{-2 \F_B} (x_+, x_-)  =- { f_B'(x_+) \tilde  f_B'(x_-) \over ( f_B(x_+) -  f_B (x_-))^2} .
\ee
The associated boundary stress tensors are
\be 
2 T(x_+) = S(f_B, x_+), \qquad 2  \tilde T(x_-) = S(\tilde f_B, x_-).
\ee

In our parametrization, working in the in the upper half plane model for $\S_0$, the bulk Liouville solutions $\F$  and $\tilde \F$ are determined by holomorphic   functions $f(z)$ and $\tilde f(\bar z )$  on the upper half plane, which take real values on the real line. Comparing the boundary stress tensors  (\ref{TLiouv}) and (\ref{TB})  in both descriptions  shows that we can identify
\be 
f_B (x_+) = f (x_+), \qquad \tilde f_B (x_-) = \tilde f ( x_- ).\label{relationFs}
\ee
We should note that this simple relation encodes a highly nonlocal map between the Liouville fields $\F, \tilde \F$ and $\F_B$. From the boundary point of point of view, the problem of finding the backreacted solution in the presence of particle sources reduces to finding real functions $f_B $ and $\tilde f_B$   which extend to multivalued functions  $f(z), \tilde f(\bar z)$  on the upper half plane which solve the monodromy problem for Liouville theory on a pseudosphere.
%, whose Schwarzian derivative satisfies
%(\ref{Schwarzeq}), where the accessory parameters  in $T(z)$ determined by requiring that all monodromies lie in $SL(2,\RR )$.
%are multivalued extensions  of the boundary functions $F (x_+) , \tilde  F (x_+)$, with appropriate branch  
\subsubsection*{Boundary gravitons and circle diffeomorphisms }
So far we have discussed how to construct   solutions corresponding to particles backreacting on the AdS vacuum solution from conformal blocks for correlators of primaries. In (2+1)D gravity, it is always to generate new solutions by `adding boundary gravitons', i.e. by performing  an improper diffeomorphism which preserves the Brown-Henneaux boundary conditions yet acts nontrivially on the boundary.   These diffeomorphisms are the classical equivalent of acting on the state with a Virasoro group element, and are parametrized by (two copies of) the group of diffeomorphisms of the circle.
As discussed in Section, these need to be included  in a complete description of the infinite-dimensional phase space,
 in casu $\calt (0,1,n) \times \calt (0,1,\tilde n)$.

In our formalism, adding boundary gravitons works as follows. We parametrize the circle by an angular coordinate $\f$, and consider a general diffeomorphism of $S^1$. The latter can be Fourier expanded as follows
\be 
\f \to \f' = \f + \sum_{n \in \ZZ} a_n e^{i n \f} , \qquad a_{-n} = \overline{ a_n} \label{diffS1}
\ee
To this diffeomorphism we associate a function $g(w)$ as
\be 
g(w) = z  e^{ i  \sum_{n \in \ZZ} a_n w^n}.
\ee
The function $g(w)$ is  holomorphic  on the unit disk and reduces to   the diffeomorphism  (\ref{diffS1})  on the boundary. Furthermore, it satisfies the reflection condition 
\be
g (1/w) = {1 \over \bar g (w) }.
\ee
From our expression (\ref{reflg}) we see that $g(w)$ is a conformal transformation on $\S_0$ (in the unit disk model) which, acting on a Liouville solution $\F$,  generates a new solution preserving the ZZ boundary condition. Similarly, we can apply a second conformal transformation of this type to the Liouville solution $\tilde \F$, and combining these  we obtain a (2+1)D gravity solution with added  boundary gravitons. 

In the simplest cases, where $n =0,1$, the incorporation of boundary gravitons  reproduces the expected phase space. Indeed, due to the symmetries of the unpunctured and once-punctured hyperbolic disk, the above diffeomorphisms lead to a family of solutions  labeled by ${\rm Diff (S^1)}/SL(2,\RR)$ for $n=0$ and ${\rm Diff (S^1)}/U(1)$ for $n=1$, in agreement with the known phase spaces $\calt (0,1,0)$ and $\calt (0,1,1)$.
It would be interesting to establish if, for two or more punctures, we similarly obtain a local parametrization of $\calt (0,1,n)$.
 
\subsection{Special cases and examples}\label{Secexs}
We conclude our study of spacetimes with a single asymptotic boundary with a discussion of some special classes of solutions and some concrete examples.
\subsubsection*{Solutions with only  chiral and anti-chiral particles}
As anticipated in section (\ref{Secsources}), the class of solutions for which our construction becomes  most tractable is that where all the particles are either chiral  or antichiral, 
i.e. $\a_i \tilde \a_i =0, \forall i$.
  In this case we
can satisfy (\ref{Vwl2}) by taking $V$ and $\tilde V$ to be as in the vacuum  AdS solution  and choosing the worldlines of the (and-)chiral  particles to be integral curves of the vector field $\pa_t + V$ (resp. $\pa t + \tilde V$). 
For example, in the upper half plane model for $\S_0$,  we take $V =- \tilde V= - \pa_z - \pa_{\bar z}$ and the (anti-) chiral particles move on
leftmoving (rightmoving) curves of constant $x_+ = {\rm Re\, } z + t$ (resp. const $x_- = {\rm Re \,} z - t$) at constant values of $y = {\rm Im\, z}$. In this case the time dependence of the solution is quasi-trivial, with the metric taking
the form (\ref{metrUHP}). The fields $\F$ and $\tilde \F$ should of course satisfy the sourced Liouville equations (\ref{Liouveqs}) and obey ZZ boundary conditions (\ref{ZZ}). %Note that the chiral insertion points $z_i$  and the anti-chiral ones $\tilde z_i$ should be distinct.
 
 Let us work out the explicit solution in the simplest case of one chiral particle and one anti-chiral one, which has not yet appeared in the literature. We take the chiral particle to be of strength $\a = 1 -a $ and to start from position $z_0$ at $t=0$, and the anti-chiral particle of strength $\tilde \a = 1 - \tilde a $  to start from position $\tilde z_0$. The conditions (\ref{ccond1}-\ref{ccond4}) fix the accessory parameters in the Liouville stress tensors, which read
 \be 
 T(z) = {1-a^2 \over 4} {(z_0-\bar z_0)^2 \over(z- z_0)^2 (z-\bar z_0)^2}, \qquad 
 \tilde T(z) = {1-\tilde a^2 \over 4} {(\tilde z_0-\bar{\tilde z}_0)^2 \over(z- \tilde z_0)^2 (z-\bar{\tilde z}_0)^2}
 \ee
 The corresponding holomorphic functions $f, \tilde f$ satisfying (\ref{Schwarzeq}) are
 \be 
 f = \left( { z- \bar z_0\over z - z_0} \right)^a, \qquad \tilde f = \left( { z- \bar{\tilde z}_0\over z - \tilde z_0} \right)^{\tilde a},
 \ee
 and  the  Liouville fields are given through (\ref{Liouvsol2}) by
 \be 
 e^{-2\F} = {a^2  |z_0- \bar z_0|^2 }  {(|z-z_0| |z-\bar z_0|)^{2 (1-a)} \over 4 \left( {\rm Im} \left( { z- \bar z_0\over z - z_0} \right)^a\right)^2}, \qquad 
 e^{-2\tilde \F} = {\tilde a^2  |\tilde z_0- \tilde{\bar z}_0|^2 }  {(|z-\tilde z_0| |z-\tilde{\bar z}_0|)^{2 (1-\tilde a)} \over 4 \left( {\rm Im} \left( { z- \tilde{\bar z}_0\over z - \tilde z_0} \right)^a\right)^2}.
 \ee
 We should note that we have implicitly  assumed that the particles move in  different 
 planes of constant ${\rm Im} z$, i.e. 
 ${\rm Im}  z_0 \neq {\rm Im}  \tilde z_0  $. Otherwise  they pass through each other at some time in the future or past, which produces a singularity that we will not analyze here.

\subsubsection*{Perturbative  solution for a second Liouville source}
The above explicit example required solving   Liouville's equation with a single delta-function source under ZZ boundary conditions.  For the case of two or more chiral (or non-chiral) particles, we would need to know the Liouville solution with two or more sources, or equivalently a classical  sphere block with at least four insertions. Since this function is  not known in closed form we will, following  \cite{Menotti:2006gc}, construct a perturbative solution in the regime that one of the sources  is much weaker than the other. We summarize the computation of  \cite{Menotti:2006gc} (see also \cite{Hulik:2016ifr}) here, which will allow us to perform an important check of our derived relation (\ref{cderF}) to vacuum blocks on the sphere. 

It is convenient to work in the Poincar\'e disk coordinate $w$, see (\ref{Cayley}). We choose to place the heavier particle with strength $\a := 1- a$ in  $w=0$, and the lighter one of dimension $\e\ll 1$ at $w = r$ with $r$ real. We expand the Liouville stress tensor as
\bea
T  &=& T_0 + \e T_1 + \calo (\e^2 )\\
T_0 &=& {1-a^2 \over 4 z^2}\\
T_1 &=& {1\over (z-r)^2 }+{1\over (z-1/r)^2} + {d_0 \over z} + {d_r \over z-r} + {d_{1\over r} \over z-1/r}.\label{accesspar}
\eea
The conditions   on the accessory parameters (\ref{ccondD1}-\ref{ccondD3}) can be used to solve for for $d_0$ and  $d_{1\over r}$, leading to
\be
T_1 = {\left(r - {1\over r}\right)^2 \over (z-r)^2\left( z-{1\over r}\right)^2} + {2r - d_r (1-r^2)\over z(z-r)\left( z-{1\over r}\right)},
\ee
and in addition they imply that $d_r$ is real. 

One then proceeds to solve the ordinary differential equation (\ref{ODET}) to first order in $\e$. We are interested in the monodromy $\d M$ picked up when encircling the point $z=r$; this can be shown to be \cite{Menotti:2006gc}
\be
\delta M_i^j = 2\p i \epsilon^{jk} \left( d_r \psi^0_i (r) \psi^0_k (r) + \psi^{0\prime}_i(r)  \psi^0_k(r) + \psi^0_i(r) \psi^{0\prime}_k(r) \right),\label{Mmon}
\ee
where $\psi^0_i$ are the zeroth-order solutions
\bea 
 \psi_1^0 & = & \frac{1}{\sqrt{a(1-|z_0|^2)}} (z-z_0)^{\frac{1+a}{2}} (1-\bar{z}_0 z)^{\frac{1-a}{2}} \\
\psi_2^0 & = & \frac{1}{\sqrt{a(1-|z_0|^2)}} (z-z_0)^{\frac{1-a}{2}} (1-\bar{z}_0 z)^{\frac{1+a}{2}}.
\label{psi0}
\eea   
Evaluating (\ref{Mmon}) gives
\bea
\d M_1^1 &=& - \d M_2^2 = {2\p i \over a} (1 + r d_r)\\
\d M_1^2 &=&  {-2\p i r^a \over a} (a+1+ r d_r) \\
\d M_2^1 &=&  {-2\p i r^{-a} \over a} (a-1- r d_r).
\eea
To obtain a single-valued Liouville field we should impose that $\d M$ belongs to $ SU(1,1) \subset SL(2,\CC)$. This leads to 
\bea
\d M_i^{\ i} &=& 0\\
\d M_2^2 &=& \overline{\d M_1^1}\\
\d M_2^1 &=& \overline{\d M_1^2}.
\eea
The first and second conditions are  automatically satisfied (recalling that $d_r$ is real), while the third one determines $d_r$ to be
\be d_r =-{1\over r} \left( 1+a {r^a + r^{-a}\over r^a - r^{-a}}\right).\label{crfrommon}
\ee

We can now verify that our main identity (\ref{cderF}) relating the Liouville accessory parameters to classical blocks  holds in this example. The classical four-point block was calculated in the same perturbative approximation in \cite{Hijano:2015rla}:
\be 
F_0(x) = \left( \ln x + 2 \ln {x^{-{a \over 2}} - x^{a \over 2} \over a} \right)\e
\ee 
where $x$ is the crossratio
\be 
x = {(z_1-z_2)(z_3-z_4) \over (z_1-z_3)(z_2-z_4)}.
\ee
For the configuration of interest, the insertion points are
\be 
z_1 =\infty,\qquad z_2 = {1\over r}, \qquad z_3 = r,\qquad z_4 =0,\label{diskconfig}
\ee 
and the crossratio is $x=r^2$. The accessory parameter corresponding to the  $z_3$ insertion is 
\bea
c_3 &=& F'(x) {\pa x \over \pa z_3}\\
&=& \left( 1 + a {x^{{a \over 2}} + x^{-{a \over 2}}\over  x^{{a \over 2}} - x^{-{a \over 2}}}\right) \left({1 \over z_1- z_3}+ {1 \over z_3- z_4}\right) \e.
\eea
In particular, for the configuration (\ref{diskconfig})  we find \be c_3 = d_r \e\ee with $d_r$ given in (\ref{crfrommon}). This confirms our basic property (\ref{cderF}).

\subsubsection*{A scaling limit}
The 2D hyperbolic metrics $ds^2_2$ and $d\tilde s^2_2$ that we introduced in Section \ref{Sechyp} are  generically auxiliary objects which are not embedded in the (2+1)D geometry in any simple way (see (\ref{3Dmetrgen})). There is however a certain scaling limit in which the (2+1)D geometry becomes a fibration over a hyperbolic base manifold with metric   $ds^2_2$.
As we shall presently see, the limit corresponds to zooming in on a small region of $d\tilde s^2_2$ such that the geometry becomes approximately flat.  This   makes contact with \cite{Hulik:2016ifr,Hulik:2018dpl} where such limiting solutions were explored in detail.

In order to zoom in on an approximately flat (with possible conical singularities from the sources) region of  $d\tilde s^2_2$, we perform a scaling limit in which the potential term in the Liouville equation can be neglected. For this purpose we make a field redefinition
\be 
\tilde \F \to \tilde \F ' + \L,
\ee
and then taking the $\L \to \infty$ limit while keeping $\F '$ and all other variables fixed. The new field satisfies Poisson's equation
\be  \pa_z \pa_{\bar z} \tilde \F'  = \p \sum_{ i=1}^{ n}  \tilde \a_i \d^{(2)}(z - z_{\tilde i}).\label{throateqs}
\ee
and the metric ${d\tilde s^2_2}'= e^{- 2 \F '} $ is locally flat. For solutions containing only chiral and antichiral particles, 
performing the scaling limit in the (2+1)D metric
(\ref{metrUHP}) leads to
\be
ds^2 
=  e^{- 2 \F(z_+, \bar z_+)} d z_+  d\bar z_+  -  \left[ {\rm Im} \left( \pa_{z_+} \F(z_+, \bar z_+) d z_+ - \pa_{z_-}\tilde\F '(z_-, \bar z_-) d z_-\right)\right]^2  ,
\ee
One can verify that this satisfies Einstein's equations. Note that attempting to take a similar scaling limit on  both $\F$ and $\tilde \F$ would lead to  a  degenerate metric. 

In the absence of antichiral sources, $\tilde \a_{ i} =0 \ \forall i$, we can take the following  solution for $\tilde \F$: 
\be 
 \tilde \F' (z, \bar z) = {\rm Im}\ z.
 \ee 
The metric then becomes
\be 
ds^2 =  e^{- 2 \F(z_+, \bar z_+)} d z_+  d\bar z_+  -  \left( dt - {\rm Im} \left( \left( { i\over 2} + \pa_{z_+} \F(z_+, \bar z_+)\right) d z_+\right) \right)^2 . 
\ee
In this limit, the  (2+1)D geometry  takes the form of a timelike fibration over a hyperbolic base manifold with metric $ds^2_2$.
This particular scaling limit for solutions with chiral particles was  studied extensively in \cite{Hulik:2016ifr,Hulik:2018dpl}. We note that, as $\tilde A_\t \sim L_0$, our criterion (\ref{chiralgeod}) is obeyed and the chiral particles move on geodesics. This was also checked explicitly in \cite{Hulik:2016ifr}.

\section{Towards holography for closed universes}\label{SecS2}
%Our formulation of (2+1)D gravity in Section \ref{} constructs multi-centered solutions  by solving two  decoupled inhomogeneous Liouville equations  (\ref{Liouveqs}) on an initial slice $\S_0$. 
So far we have considered a time slice $\S_0$ with an asymptotic boundary, as appropriate for describing particles in an asymptotically AdS spacetime. However our formalism applies in principle to  time slices of arbitrary topology, 
and it is interesting to consider the case where  $\S_0$ is a compact Riemann surface without boundary. The  corresponding (2+1)D solutions are closed  universes evolving from a big bang to big a crunch singularity and were considered in a related context in \cite{Mess,Scarinci:2011np}, see also \cite{Carlip:2004ba} for a review and further references. The study of such boundary-less slices generalizes the standard AdS/CFT setup % with conformal boundaries
and could be a step towards studying cosmological singularities. 

If  $\S_0$ is a surface of genus $g$, the Gauss-Bonnet theorem places a necessary condition on the parameters  $0\leq \a_i < 1$ for a solution to exist \cite{Seiberg:1990eb}:
\be \sum_{i=1}^n \a_i > 2 (1-g) .\label{Seibergbound}\ee
For simplicity we assume in what follows that $\S_0$  has spherical topology, $g=0$. Note that we need $n \geq 3$ to satisfy (\ref{Seibergbound}). The classical result going back to Picard \cite{Picard} is that, when  (\ref{Seibergbound}) is obeyed, a solution to the inhomogeneous Liouville equation (\ref{Liouveqssource}) exists and is unique. 

As we did in the case with asymptotic boundary, we will study a semiclassical Hartle-Hawking path integral preparing  the multi-particle state on $\S_0$, which will again be closely related to the Liouville action on $\S_0$. We  then go on to examine the role of classical Virasoro blocks in determining this wavefunction and the   gravity solution.

\subsection{Hartle-Hawking wavefunction and Liouville  action}
%In this subsection we will establish a relation between the Hartle-Hawking wavefunction
 We are interested in computing a  Chern-Simons path integral, which is analogous to  the Hartle-Hawking \cite{Hartle:1983ai} path integral over metrics, and prepares a multi-particle state  on the Riemann sphere $\S_0=\overline{\CC}$ at $t=0$.  
 We take the  particles to be located at  $z_i , i = 1, \ldots, n$ with quantum numbers $\a_i, \tilde \a_i$.
 By applying  M\"obius isometries of $\S_0$  we can fix the last three particle locations to 
\be 
z_{n-2} = 0, \qquad z_{n-1} = 1, \qquad z_{n} = \infty.
\ee
Near the sources, the Liouville field $\F$ has the asymptotics
\bea 
\F &\ \ \ \ \stackrel{\footnotesize\mathclap{\mbox{$z \to z_j$}}}{\sim}\ \ \ \ &  \a_j \ln |z -z_j|+\calo (1) , \qquad j = 1, \ldots, n-1 \label{Phinearpunct}\\
 &\ \ \ \ \stackrel{\footnotesize\mathclap{\mbox{$z \to \infty$}}}{\sim}\ \ \ \  & \left(2 - {\a_n } \right) \ln |z|+\calo (1), \label{Phinearinfty}
\eea
and similarly for $\tilde \F$. 
%We continue to imaginary time, $t=i \t$ \comment{Maybe this is not necessary in this topological theory?} and

In analogy with the Hartle-Hawking no-boundary proposal we want to   perform a path integral on  a 3-manifold $X$ whose boundary at $t=0$ is $\S_0$ and without boundaries in the past. Each particle location $z_i$  is the endpoint of a worldline in $X$. Since our framework is not equipped to deal with worldlines ending on each other\footnote{To describe more general configurations, we need a framework where worldlines can end on each other, in other words  which incorporated  bulk interactions. This is beyond the scope of the present work, see however \cite{Hijano:2015qja}.} - for one thing, our gauge choice would break down at such an endpoint -  each worldline needs to connect two particle locations on $\S_0$.  We therefore restrict attention to configurations of particles at $\S_0$  which come in  pairs  with the same quantum numbers $\a_i, \tilde \a_i$. The manifold $X$ is then a ball containing particle worldlines which connect pairs of boundary points (see Figure \ref{FigS2radialtime}). 
%\begin{figure}
%	\begin{center}
%		\includegraphics[height=150pt]{S2radialtime.pdf}
%		\end{center}
%	\caption{}
%	\label{FigS2radialtime}
%\end{figure} 
 \begin{figure}
%	\begin{center}
	\begin{picture}(50,200)
	\put(130,-20){\includegraphics[height=180pt]{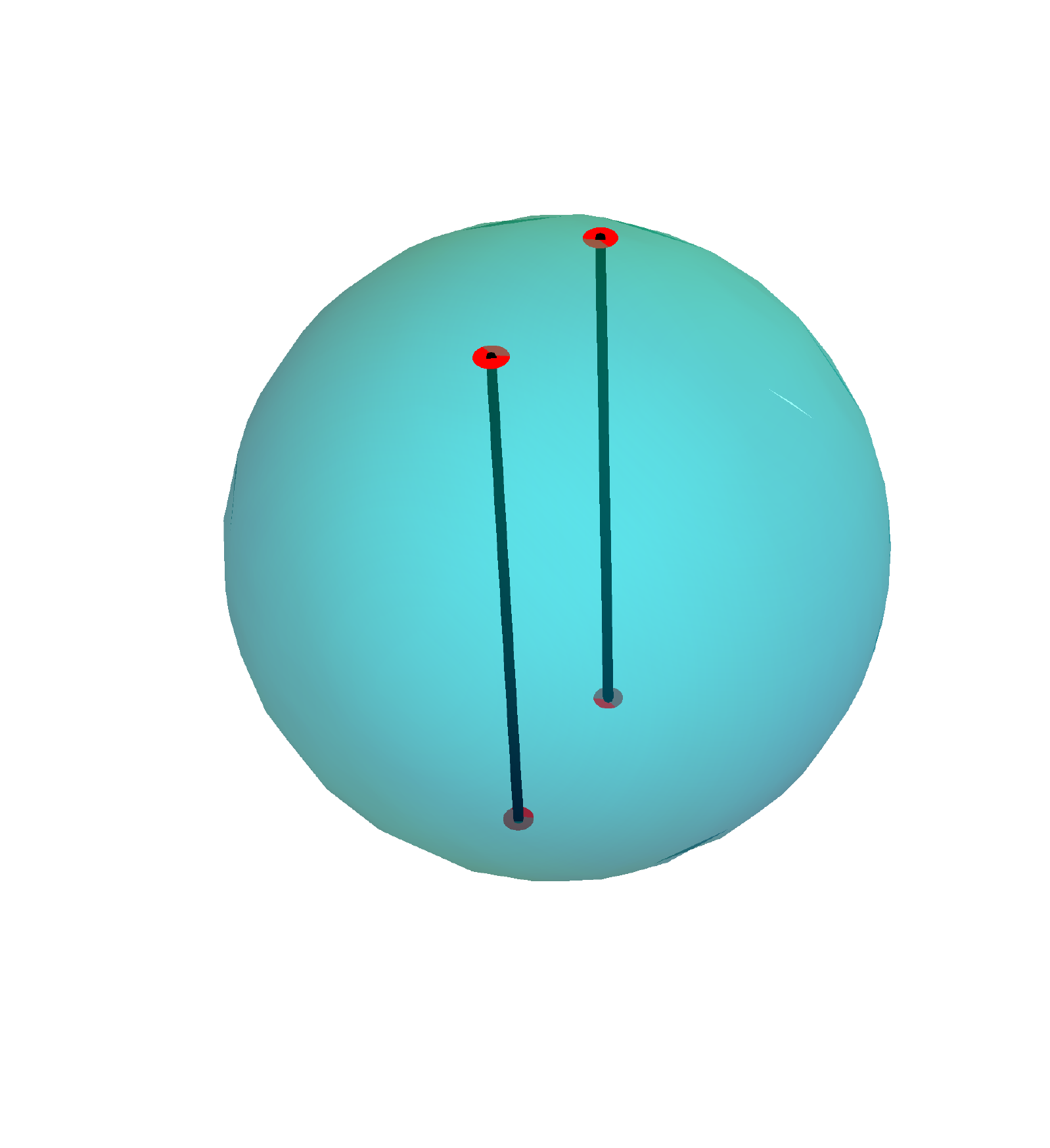}}
	\end{picture}
%		\end{center}
	\caption{The lower half of Figure \ref{FigHH}(c) can be redrawn as a ball  with particle worldlines connecting  points on the boundary two-sphere through the bulk.}
	\label{FigS2radialtime}
\end{figure}

Proceeding as in Section \ref{SecHH}, we find that the total action evaluates to a boundary term which reads
 \be 
i S_{CS}^{tot} = -  {k \over 4} S_L[ \F] + {3 i k\over 8 \p }\lim_{\e \to 0} \int_{\overline{\CC}_\e}  dz \wedge d \bar z  e^{-2 \F},
 \ee
where  $S_L [\F]$ is the  standard, regularized, Liouville action \cite{Zamolodchikov:1995aa}:
  \bea 
 S_{L}[\F] &=& \lim_{\e \to 0}  S_{L}^\e[\F]\nonu
 S_{L}^\e[\F] &=&  {i \over 2 \p} \left(  \int_{\overline{\CC}_\e} dz \wedge d \bar z \left( |\pa_z \F|^2 + e^{-2 \F} \right)+ r_\e\right. \nonu
 && - \sum_{j=1}^{n-1}{\a_j \over 2} \oint_{C_j^\e} \F \left( {d\bar z\over \bar z - \bar z_j} - {d z\over  z -  z_j} \right)\nonu
 && \left.- \left(1 - {\a_n\over 2}\right)
 \oint_{C_n^\e} \F \left( {d\bar z\over \bar z } - {d z\over  z  } \right)\right). \label{SLS2}
 \eea 
Here, $\overline{\CC}_\e$ 
is the extended complex plane with disks of radius $\e$ removed around the insertion points $z_i$ (including $z_n = \infty$)
%the insertion points $z_i, i = 1, \ldots n-1$  
and $C_i^\e$ are the corresponding boundary curves. The constant
  \be 
 r_\e = - \p \left( \sum_{j=1}^{n-1} \a_j^2 + ( 2- \a_n)^2\right) \ln \e.
 \ee
 is included to obtain a finite result in the $\e\to 0$ limit.
 
 Using the fact the source-free Liouville equation holds on $\overline{\CC}_\e$ and the asymptotic behavior (\ref{Phinearpunct},\ref{Phinearinfty}), the last term can be evaluated to yield
  \be 
i S_{CS}^{tot} =-  {k \over 4} S_L[ \F] - {3  k\over 8  }\left(\sum_{j-1}^n \a_j - 2\right) .\label{SCSSLS2}
 \ee
Therefore, the on-shell Chern-Simons action is once again proportional to the on-shell Liouville action, up to an additive constant  depending only on the quantum numbers $\a_i$ of the particles but not on their locations $z_i$. 

\subsection{Liouville action,  accessory parameters and classical blocks}

Let us review the Polyakov relations satisfied by the on-shell Liouville action, as well as its interpretation in terms of the geometry of Teichm\"uller space.
 As in the case with conformal boundary, the problem of solving  the sourced  Liouville equation can be recast as a monodromy problem for a set of accessory parameters.
The 
Liouville stress tensor (\ref{TLiouv}) is a meromorphic function  of the form
\be
T (z) =  \sum_{i=1}^{n-1}\left( {\e_i \over  (z-z_i)^2}+ {c_i \over z-z_i}\right),\label{TS2}
\ee
%where
%\be 
%\e_i = {\a_i \over 2} \left( 1- {\a_i \over 2 } \right)%{\d_i \over 4 \p} \left( 2- {\d_i \over 2 \p} \right) %=  2 {h_i \over k}\left( 1-{h_i \over k}\right).
%\ee
In addition, from (\ref{Phinearinfty}) the behavior of $T$ near $z\to \infty$ is given by
\be 
T (z)\sim  {\e_n \over z^2} + {c_n \over z^3}+ \calo ( z^{-4}) , \qquad z\to  \infty.\label{Tinfty}
\ee

The   accessory parameters $c_i$ 
are subject to three three linear relations imposed by  the asymptotics (\ref{Tinfty}),
\be 
\sum_{i=1}^{n-1} c_i =0, \qquad \sum_{i=1}^{n-1} ( \e_i + c_i z_i) = \e_n, \qquad \sum_{i=1}^{n-1} (2 \e_i  z_i + c_i z_i^2) = c_n.
\ee
Therefore the independent accessory parameters can be taken to be $c_1, \ldots , c_{n-3}$.

The  Liouville solution is again obtained from solutions to the ODE (\ref{ODET}), whose monodromies around the singular points  should once again lie in the subgroup $SL(2\,\RR) $
of $SL(2,\CC)$. One can show (see e.g. \cite{Hulik:2016ifr}) that this imposes $2 (n-3)$ real conditions, precisely  as many  as there are undetermined accessory parameters . The existence and uniqueness of the Liouville solution implies  that this monodromy problem indeed uniquely fixes the  accessory parameters.

The accessory parameters have an important  interpretation in terms of the K\"ahler geometry of Teichm\"{u}ller space. It can be shown \cite{Zograf_1988} that they are real-analytic
functions of the moduli. They are however not holomorphic, and their antiholomorphic derivatives give the Weil-Petersson metric on Teichm\"{u}ller space:
\be 
{\pa c_i \over \pa \bar z_j} = g^{WP}_{i \bar j}. \label{cWP}
\ee
A conjecture by Polyakov, which was rigorously proven in \cite{Zograf_1988,Takhtajan:2001uj}, states that the   on-shell Liouville action (\ref{SLS2}) is a generating function of the accessory parameters:
\be 
{\pa S_L \over \pa  z_i} = c_i. \label{SLcS2}
\ee
Combining this with (\ref{cWP}) shows that $S_L$ is a K\"ahler potential for the Weil-Petersson metric.  Due to the property (\ref{SCSSLS2}), this is also the case for  the total action  $i S_{CS}^{tot}$.

A derivation of Polyakov's conjecture (\ref{SLcS2}) along the lines of \cite{Takhtajan:2001uj}  proceeds as in Section \ref{Secaccess}.  We want to compute
\be 
\lim_{\e \to 0} \pa_{z_i} S_L^\e,
\ee
with $S_L^\e$ given in (\ref{SLS2}). The following identity takes into account the variations of the integration domain $\overline{\CC}_\e$
\be 
\pa_{z_i} \int_{\overline{\CC}_\e} dz \wedge d \bar z \, G = \oint_{C_i^\e} G\, d\bar z + \int_{\overline{\CC}_\e} dz \wedge d \bar z\, \pa_{z_i}  G.\label{dersdomainS2}
\ee
Making use of (\ref{dersdomain}) we find
\bea
 \pa_{z_i} S_L^\e &=&  {i \over 2 \p} \left( \oint_{C_i^\e} \left( |\pa_z \F|^2 + e^{-2 \F} \right)  d\bar z + \sum_{j=1}^n   \oint_{C_j^\e} \pa_{z_i}\F \left(\pa_{\bar z}\F d\bar z - \pa_z\F dz\right)\right.\nonu
 && - \half \sum_{j=1}^{n-1} \a_j  \oint_{C_j^\e}
 \left( \pa_{z_i} \F + \d_{ij} \pa_z \F \right) \left( {d\bar z\over \bar z - \bar z_j} - {d z\over  z -  z_j} \right)\nonu
 && \left.- \left(1 - {\a_n\over 2}\right)
 \oint_{C_n^\e}\pa_{z_i} \F \left( {d\bar z\over \bar z } - {d z\over  z  } \right) \right),\label{derSLS2}
 \eea
 where each line in this expression is the derivative of the corresponding line in (\ref{SLS2}).
 To evaluate this in the $\e \to 0$ limit  we  need the first subleading terms in the expansions
  (\ref{Phinearpunct}, \ref{Phinearinfty}). The form of the stress tensor (\ref{TS2}) determines these to be of the form  
\bea 
\F &\ \ \ \ \stackrel{\footnotesize\mathclap{\mbox{$z \to z_j$}}}{\sim}\ \ \ \ &  \a_j \ln |z -z_j|+ \s_j - {c_j \over \a_j}(z-z_j) - {\bar c_j \over \a_j}(\bar z-\bar z_j) +\ldots, \qquad j = 1, \ldots, n-1 \label{Phinearpunct2}\nonu
 &\ \ \ \ \stackrel{\footnotesize\mathclap{\mbox{$z \to \infty$}}}{\sim}\ \ \ \  & \left(2 - {\a_n } \right) \ln |z|+\s_n -{c_n \over  \a_n}{1\over z}- {\bar c_n \over  \a_n}
 {1\over \bar z}+\ldots,\label{Phinearinfty2}
\eea
where the $\s_i$ are functions of $(z_j, \bar z_j )$.  Using these expansions in (\ref{derSLS2})   and taking the $\e \to 0$ limit we obtain 
\bea 
\pa_{z_i} S_L &=& \sum_{j=1}^{n-1} \a_j \pa_{z_i} \s_j + (\a_n- 2) \pa_{z_i} \s_n + c_i\nonu
&&  - \sum_{j=1}^{n-1} \a_j \pa_{z_i} \s_j\nonu
&& -  (\a_n- 2) \pa_{z_i} \s_n.
\eea
%-  \sum_{j=1}^n \a_j \s_j + 2 \s_n  .
Adding these up we arrive at (\ref{SLcS2}).

Having reviewed the Polyakov relation (\ref{SLcS2}) we can now relate the accessory parameters $c_i$ to the classical Virasoro blocks.  The unique solution to the inhomogeneous Liouville equation determines a stress tensor  $T(z)$ whose associated ODE, in some chosen OPE channel, has monodromies corresponding to a specific set of exchanged families $\n_{i'}^*$. Therefore the  $c_i$ should   coincide with the accessory parameters  in the monodromy problem determining a  classical block in the  chosen channel. In the notation of section (\ref{Secaccess}) we have
\be 
c_i (z_j, \bar z_j ) = D_i (z_j, \e_i, \n_{i'}^*) .\label{cisC}
\ee
We note that, for consistency with (\ref{cWP}), the dimensions  $\n_{i'}^*$ must implicitly depend on the $z_j$ and their complex conjugates $\bar z_j$. 
We can say a bit more on this dependence from the known properties of Liouville theory. It can be argued from the large-$k$ behavior of  of  Liouville correlators that the on-shell action is of the form \cite{Zamolodchikov:1995aa}
\be 
S_L =  \left.\left( \cals^{(3)} (\e_i, \n_{i'} ) + F(z_i, \e_i, \n_{i'} ) +  \bar F(\bar z_i, \e_i, \n_{i'} ) \right)\right|_{\n_{i'}=\n_{i'}^*}
\ee
Here, $\cals^{(3)}$ contains  the  contribution from   three-point coefficients at large-$k$, and the right-hand side is evaluated on exchanged dimensions $\n_{i'}^* (z_j, \bar z_j)$ which  extremize this contribution.
This property, together with the Polyakov relations (\ref{SLcS2}),  then implies  the relation (\ref{cisC}), since
\be 
{\pa S_L \over \pa z_i} = {\pa F ( z_i, \e_i, \n_{i'}^*)  \over \pa z_i}.
\ee

To conclude, we have established that the  Hartle-Hawking wavefunction for a closed universe containing some point particles is once again determined by the on-shell Liouville action. In contrast to the situation with asymptotic boundary however, the action and  accessory parameters contain dynamical information of Liouville theory beyond the kinematical data contained in  conformal blocks. This information enters through the 
nontrivial dependence of the exchanged families on the insertion points $\n_{i'}^* (z_j, \bar z_j)$.

To emphasize the increased challenge associated to the closed surface situation, it is helpful to compare how the Liouville problem on a closed surface and on the pseudosphere problem relate to the conformal block problem of section \ref{blockproblem}. The Liouville monodromy problem requires us to restrict the monodromies to fall within $SL(2,\mathbb{R})$, while in the conformal block problem one fixes a smaller number of monodromies but in addition one fixes the exact conjugacy classes of $SL(2,\mathbb{R})$. In the case of the pseudosphere these problems coincide due to the additional structure provided by the reflection symmetry. This fixes the resulting conformal block to the identity block of the channel depicted in figure \ref{fig:OPEchannel}. 

In the closed surface case, where there is no such additional structure, the solution to the Liouville problem will once again provide a conformal block, but which conformal block one obtains is a non-kinematical question that is sensitive to the OPE structure of Liouville theory.

\section{Outlook}
We conclude  by listing some open problems and future directions.
\begin{itemize}
\item In this work we restricted attention to spacetimes containing point particles but no black holes. It is of obvious interest to generalize our discussion to bulk states containing black holes. A first  interesting configuration  would be to consider states containing  point particle excitations on top of an eternal BTZ black hole (see also \cite{Jackson:2014nla}). In this case the slice $\S$ would contain a second asymptotic boundary. A second question would be to address dynamical black hole formation from the collapse of point particles \cite{Matschull:1998rv}. We would for example hope that  our framework could make more explicit  make the link between conformal blocks with a continuous  operator distribution  and Vaidya-type metrics  proposed in \cite{Anous:2016kss}.
\item  Since the considerations in this work were semiclassical, it would be of great interest to extend them to the quantum regime. One of the difficulties in doing so is the quotient by the mapping class group in the gravitational phase space (see Footnote \ref{FNMappingclass}),   which is hard to implement in the Chern-Simons path integral as pointed out in \cite{Eberhardt:2022wlc}. It would also be useful to clarify how our semiclassical wavefunctions emerge at large $k$ from those considered in \cite{Verlinde:1989ua}, which were derived in the  `quantize first and then constrain' approach.
\end{itemize}

 \section*{Acknowledgements}
JR would like to thank Ondra Hul\'ik and Orestis Vasilakis for initial collaboration and valuable discussions. We furthermore thank T. Anous, S. Banerjee, M. Kudrna and T. Proch\'azka for useful discussions. This work was supported
by the Grant Agency of the Czech Republic under the grant EXPRO 20-25775X.
\begin{appendix}
\section{MPD equations}\label{AppMPD}
The Mathisson-Papapetrou-Dixon (MPD) equations describe the motion of an intrinsically spinning particle in general relativity. They take the form \cite{Papapetrou:1951pa}
\bea 
\nabla_s (m u^\m + u_\n \nabla_s S^{\m\n} )&=& \half R^\m_{\ \n\r\s} u^\n S^{\r \s}\label{MPD1}\\
\nabla_s S^{\m\n} + u^\m u_\r \nabla_s S^{\n\r}-  u^\n u_\r \nabla_s S^{\m\r} &=&0\label{MPD2}
\eea
where  $S^{\m\n}$ is an antisymmetric tensor called the spin tensor and $u^\m = {d x^\m \over ds}$ is the velocity of the worldline parametrized by proper time, {$ u^\m u_\m =-1$.} The worldline covariant derivative is defined as
\be 
\nabla_s v^\m = {d v^\m \over ds} + \G^\m_{\n\r} u^\n v^\r.
\ee

We are interested in the MPD equations in (2+1)D  spacetimes which are locally AdS. The right-hand side of (\ref{MPD1})  simplifies in this case to
\be
\half R^\m_{\ \n\r\s} u^\n S^{\r \s} =  S^{\m\n} u_\n .
\ee
%In 3D the most general form the spin tensor can take is $S^{\mu\nu} = \epsilon^{\mu\nu\rho}k_{\rho}$.
The MPD equations  then allow for special class of solutions where the worldline  is a geodesic, i.e.
\bea 
%S^{ \m\n} &=& \s \e^{\m\n\r} u_\r\\
\nabla_s u^\m &=& 0,\label{geodsol1}
\eea
One can  show that (\ref{MPD1},\ref{MPD2}) then  imply that
\be
S^{ \m\n} = \s \e^{\m\n\r} u_\r,\label{geodsol2}
\ee
%were $\s$ is the spin of the particle. \comment{Is $\s = |s|$ defined in main text?}
where $\s$ is an arbitrary proportionality constant. 

Following the steps in Appendix E of \cite{Castro:2014tta}, the MPD equations (\ref{MPD1},\ref{MPD2}) can be rewritten as
\bea
{d P \over d s} + [ \O_s + E_s, P] &=&0\label{MPDP1}\\
{d \tilde P \over d s} + [ \O_s - E_s,\tilde P] &=&0\label{MPDP2}
\eea
where $\O_s = \half \e^a_{\ bc} \O_{\m bc} u^\m J_a,\ E_s = E_\m^a u^\m J_a$, $E$ and $\O$ are the (2+1)D vielbein and spin connection,  and $J_a$ are $sl(2,\RR)$ generators satisfying $[J_a, J_b] = \e_{abc} \h^{cd} J_d$. The quantitites $P$ and $\tilde P$ in (\ref{MPDP1},\ref{MPDP2}) are defined as
\bea
P := \left( m u^a + u_b \nabla_s S^{ab} + \half \e^a_{\ bc} S^{b c}\right) J_ a\\
\tilde P := \left( m u^a + u_b \nabla_s S^{ab} - \half \e^a_{\ bc} S^{b c}\right) J_ a.
%P := \half \left( m u^a + u_b \nabla_s S^{ab} + \half \e^a_{\ bc} S^{b c}\right) J_ a\\
%\tilde P := \half \left( m u^a + u_b \nabla_s S^{ab} - \half \e^a_{\ bc} S^{b c}\right) J_ a.
\eea
The form (\ref{MPDP1},\ref{MPDP2}) of the MPD equations  precisely coincides with the equations of motion  (\ref{CSmat1}) for the Chern-Simons-matter system in the main text. Note that (\ref{MPDP1},\ref{MPDP2}) imply that $\tr P^2 $ and $\tr \tilde P^2 $ are constant, so that the additional  equation of motion (\ref{CSmat2}) serves to fix integration constants in terms of the mass and spin of the particle.

For example, for solutions describing geodesic motion (\ref{geodsol1},\ref{geodsol2}), the momenta take the form
\be 
P = {(m + \s) } u^a J_a, \qquad P = {(m - \s )} u^a J_a\label{geodsolP}
\ee
The equation (\ref{CSmat2}) then tells us that $\s$ is to be identified with the particle helicity $s$.  From (\ref{geodsol1}, \ref{geodsolP}) we infer that general solutions describing geodesic motion are characterized by
\be 
 [P, \tilde P] = \nabla_s P = \nabla_s \tilde P =0,
 \ee
 or, equivalently, making use of (\ref{MPDP1},\ref{MPDP2}),
 \be 
 [P, \tilde P] =  [E_s, P] = [E_s, \tilde P] =0. \label{geodsubclassPApp}
 \ee

\section{Some formulas for the Poincar\'e disk model}\label{Appdisk}
In this Appendix we collect some formulas relevant when using the Poincar\'e disk model for $\S_0$. As in (\ref{Cayley}) we use a complex coordinate $w, \ |w|<1$.

Let us start  by deriving the form of the (2+1)D metric.
A natural choice for the vector fields $V, \tilde V$ specifying the Chern-Simons gauge  choice (\ref{covgauge}) is
\be 
V =  - i w \pa_w + i \bar w \pa_{\bar w}, \qquad  \tilde V =  i w \pa_w - i \bar w \pa_{\bar w}.
\ee
From our time evolution equations (\ref{dyneqs}) we see that the fields $\l, \tilde \l$ should be turned on at times $t \neq 0$, leading to
\begin{align}
\F (t ,w, \bar w) &=  \F ( w e^{it}, \bar w e^{- it} ), & \tilde \F (t ,w, \bar w) &=  \F ( w e^{-it}, \bar w e^{ it} )\\
\l &= - t, & \tilde \l & = t\\
\m &=0,&  \tilde \m &=0 
\end{align}
Defining the combinations
\be 
w_+= w e^{ i t}, \qquad w_=  w e^{- i t},
\ee
corresponding 3D metric is 
\be
ds^2 
= \left| e^{- \F(w_+, \bar w_+)} d z_+   +  e^{-  \tilde  \F(w_-, \bar z_-)} d w_- \right|^2 -  \left[{\rm Im} \left( \pa_{w_+} \F(w_+, \bar w_+) d w_+ - \pa_{w_-}\tilde\F(w_-, \bar w_-) d w_-\right)\right]^2 . \label{metrD}
\ee

We now summarize  some formulas needed for solving the  inhomogeneous Liouville equation on the unit disk with ZZ boundary conditions, referring to \cite{Hulik:2016ifr} for more details.  Asymptotically AdS solutions are described by Liouville fields with ZZ boundary conditions which take the form
\be
e^{2 \F} = { (1-|w|^2)^2} +\calo( (1-|w|^2)^4 ),
\ee
and the stress tensor obeys
 \be
 \left. \left( {w^2} T (w)\right)\right \rvert_{|w|=1} \in \RR.\label{realTL}
 \ee
The doubling trick extending $T$ to the Riemann sphere is
\be
T (w) = {1 \over w^4} \bar T  (1/w).\label{reflD}
\ee
 A general Liouville solution is specified by a function $g(w) = \psi_1(w)/\psi_2(w)$ through
 	\be
	e^{-2 \F}=(\psi_1 \bar \psi_1 -\psi_2 \bar \psi_2)^{-2}  =  {|g'|^2 \over (1- |g|^2)^2},\label{Phig}
	\ee
 and the appropriate doubling trick for $g$ reads
\be
g(w)= {1\over \overline{ g} (1/ w)}.\label{reflg}
\ee

Now let us consider  the presence of $n$ particle sources in locations $w_i$, where  we assume\footnote{See \cite{Hulik:2016ifr} for a discussion of the general case.} for definiteness that $w_n=0$. 
The stress tensor of an inhomogeneous Liouville solution is of the form
\be
T(w) = {\epsilon_n \over w^2} + {c_n\over w} + \sum_{i=1}^{n-1}  \left({\epsilon_i \over  (w-w_i)^2}+ { \tilde \epsilon_i \over  (w-1/\bar w_i)^2}+ {c_i\over w-w_i}+ {\tilde c_i\over w-1/\bar w_i}\right),
\ee
where  the accessory parameters $c_i, \tilde c_i$ are constrained by regularity at infinity and by the reflection condition (\ref{reflD}) to obey  
\bea
2 \epsilon_i + c_i w_i + {\bar{\tilde c}_i \over w_i} &=& 0\label{ccondD1}\\
c_n + \sum_{i=1}^{n-1} (c_i + \tilde c_i) &=&0\label{ccondD2}\\
{\rm Im\,} \left( \sum_{i=1}^{n-1}  \left(c_i w_i  - {\bar{\tilde c}_i\over w_i} \right)\right)&=&-  2 \e_n \label{ccondD3}.
\eea
The accessory parameters are determined by requiring the monodromy of the ODE (\ref{ODET}) around each of the sources to lie in $SU(1,1) \subset SL(2,\CC)$, so that the Liouville field (\ref{Phig}) is single-valued.
\end{appendix}

\bibliographystyle{ytphys}
\bibliography{ref}
\end{document}